\documentclass[11pt]{JHEP3}
\usepackage{latexsym}
\usepackage{graphics}
\usepackage{amsmath}
\usepackage{amsfonts}

\newcommand{\bmat}{\left(\begin{array}}
\newcommand{\emat}{\end{array}\right)}
\newcommand{\beq}{\begin{equation}}
\newcommand{\eeq}{\end{equation}}
\newcommand{\sub}[1]{\phantom{}_{(#1)}\phantom{}}

\newtheorem{lemma}{Lemma}[section]

\def\yzero{\smash{\hbox{$y\kern-4pt\raise1pt\hbox{${}^\circ$}$}}}
\def\p{\partial}
\def\a{\alpha}
\def\b{\beta}
\def\g{\gamma}
\def\d{\delta}
\def\Om{\Omega}
\def\om{\omega}
\def\th{\theta}

\def\-{\hphantom{-}}

\def\s2{\frac{1}{\sqrt2}}

\def\beq{\begin{equation}}
\def\eeq{\end{equation}}

\def\be{\begin{equation}}
\def\ee{\end{equation}}

\def\beqa{\begin{eqnarray}}
\def\eeqa{\end{eqnarray}}

\def\bea{\begin{eqnarray}}
\def\eea{\end{eqnarray}}

\def\half{\frac{1}{2}}

\def\IF{\relax{\rm I\kern-.18em F}}
\def\II{\relax{\rm I\kern-.18em I}}
\def\IP{\relax{\rm I\kern-.18em P}}

\def\cn{{\cal N}}
\def\cm{{\cal M}}

\def\ch{{\cal H}}
\def\cf{{\cal F}}

\def\Dsl{\,\raise.15ex\hbox{/}\mkern-13.5mu D} 

\def\IC{\bf C}
\def\IZ{\bf Z}

\def\z2z2{$\IC^3/(\IZ_2\times\IZ_2)$}

\def\a{\alpha}
\def\b{\beta}

\def\d{\delta}
\def\e{\epsilon}
\def\f{\phi}
\def\g{\gamma}

\def\k{\kappa}
\def\l{\lambda}
\def\m{\mu}
\def\n{\nu}

\def\p{\pi}

\def\r{\rho}
\def\s{\sigma}
\def\t{\tau}
\def\x{\xi}
\def\z{\zeta}
\def\D{\Delta}
\def\F{\Phi}
\def\G{\Gamma}

\def\L{\Lambda}

\def\S{\Sigma}

\def\X{\Xi}

\def\vf{\varphi}

\def\ca{{\cal A}}

\def\cf{{\cal F}}

\def\ch{{\cal H}}

\def\cl{{\cal L}}
\def\cm{{\cal M}}
\def\cn{{\cal N}}
\def\co{{\cal O}}

\def\cq{{\cal Q}}

\def\bo{{\raise-.3ex\hbox{\large$\Box$}}}               
\def\pa{\partial}                                       
\def\face{{\raise.2ex\hbox{$\displaystyle \bigodot$}\mskip-2.2mu \llap {$\ddot
        \smile$}}}                                      

\def\Bar#1{\overline{#1}}                       
\def\leftrightarrowfill{$\mathsurround=0pt \mathord\leftarrow \mkern-6mu
        \cleaders\hbox{$\mkern-2mu \mathord- \mkern-2mu$}\hfill
        \mkern-6mu \mathord\rightarrow$}       
\def\dvec#1{\vbox{\ialign{##\crcr
        \leftrightarrowfill\crcr\noalign{\kern-1pt\nointerlineskip}
        $\hfil\displaystyle{#1}\hfil$\crcr}}}           

\def\beq{\begin{equation}}
\def\eeq{\end{equation}}

\def\beqx{\begin{displaymath}}
\def\eeqx{\end{displaymath}}

\def\beqa{\begin{eqnarray}}
\def\eeqa{\end{eqnarray}}
\def\NO{\nonumber}
\def\nonu{\nonumber \\}

\def\vf{\varphi}

\def\>{\rangle} 

\def\<{\langle} 

\title{Thermodynamics of Asymptotically Locally AdS Spacetimes}

\author{Ioannis Papadimitriou and Kostas Skenderis\\ Institute for Theoretical
Physics, 
University of Amsterdam, Valckenierstraat 65, 1018 XE Amsterdam, 
The Netherlands.\\ E-mail: \email{ipapadim@science.uva.nl}, 
\email{skenderi@science.uva.nl}}

\abstract{We formulate the variational problem for AdS gravity 
with Dirichlet boundary conditions and demonstrate  that the covariant
counterterms are necessary to make the variational problem well-posed.
The holographic charges associated with asymptotic symmetries are then
rederived via Noether's theorem and `covariant phase space' techniques.
This allows us to prove the first law of black hole mechanics for general
asymptotically locally AdS black hole spacetimes. We illustrate our 
discussion by computing the conserved charges and verifying the 
first law for the four dimensional Kerr-Newman-AdS
and the five dimensional Kerr-AdS black holes.}

\preprint{ITFA-2005-18\\
hep-th/0505190}

\begin{document}

\tableofcontents
\addtocontents{toc}{\protect\setcounter{tocdepth}{2}}

\section{Introduction}
\setcounter{equation}{0}

The study of gravitational theories with negative cosmological constant
has been an active area of research. The recent interest stems in 
part from the duality between asymptotically AdS spacetimes and 
quantum field theories residing at the conformal boundary of the spacetime.
One of the implications of the AdS/CFT duality is that such 
gravitational theories should exist under more general boundary conditions
than those considered in the past. 
 The boundary conditions that were considered in the previous 
works, see for instance \cite{AM,Henneaux:1984xu}, are such that the spacetime 
asymptotically approaches the exact AdS solution. In the AdS/CFT 
correspondence the fields parameterizing the boundary conditions
of bulk fields are interpreted as sources that couple to gauge 
invariant operators, and since such sources are in general 
arbitrary, we are led to consider spacetimes
with general boundary conditions.  In particular,
instead of considering the boundary conformal structure to be 
that of exact AdS, one can consider a general conformal structure. 
Such more general boundary conditions
have been considered in the mathematics literature \cite{FG} 
(see \cite{Graham,anderson} for reviews.)
We will call such spacetimes asymptotically locally AdS (AlAdS) spacetimes.

An important aspect of asymptotically AdS spacetimes, which has 
attracted considerable attention over the years, is the definition of 
conserved charges associated with the asymptotic symmetries of such 
spacetimes \cite{Abbott:1981ff,AM,Henneaux:1984xu,Hawking:1995fd,
Chrusciel:2001qr},
see also \cite{Hollands:2005wt} and references
therein. Besides the difficulty of defining 
precisely what one means by `asymptotically AdS spacetimes', the main 
obstruction in defining such conserved charges is the fact that the 
non-compactness of these spacetimes 
causes various `natural candidates' for conserved quantities, such
as Komar integrals, to diverge \cite{Magnon:1985sc}. 
One is then forced to introduce
some regularization procedure, which is inherently ambiguous.

To circumvent this difficulty, most approaches either require that the
spacetime approaches asymptotically the exact AdS metric and they 
then use the special properties of AdS to construct conserved quantities
which are manifestly finite, e.g. \cite{AM,Henneaux:1984xu}, or they embed 
the spacetime into a spacetime with the same asymptotics and then 
define manifestly finite conserved quantities relative to the ambient
spacetime, e.g. \cite{Deruelle, Barnich&Compere}. Although the 
philosophy and the precise definition of the conserved charges
varies among these methods, they all implement some form of
`background subtraction'. Therefore, despite the simplicity
and elegance of some of these methods, they are all ultimately
rather restrictive since not all asymptotically locally AdS
spacetimes can be embedded in a suitable ambient spacetime.

Inspired by the AdS/CFT correspondence \cite{ads/cft,GKP,Witten:1998qj}, a 
general background independent definition of the conserved charges 
for any AlAdS spacetime was developed in \cite{HS,Balasubramanian&Kraus,
dHSS,Skenderis_proceedings}. 
In a first step one associates to any AlAdS spacetime
a finite Brown-York stress energy tensor \cite{Brown:1992br}
obtained by varying the on-shell gravitational 
action w.r.t. the boundary metric. Finite conserved charges associated to
asymptotic symmetries are then obtained from this holographic stress 
energy tensor by a standard procedure. The finiteness of the on-shell action 
and thus of the holographic stress energy tensor and of all holographic
conserved charges is achieved by adding to the gravitational action 
a set of boundary covariant counterterms. This method is thus
also referred to as the  `method of covariant counterterms'. 
The method has been used extensively over the last few years. 
Nevertheless, there exists an ongoing debate about the connection between
these holographic charges and the various alternative definitions
of conserved charges and, in particular, regarding the validity of the
first law of black hole mechanics \cite{GPP}. One of the aims of this
paper to clarify the concept of the holographic charges (see also 
\cite{Hollands:2005ya}) and, in particular,
to prove in general that {\it all} AlAdS black holes satisfy 
the first law of
black hole mechanics and the charges entering this law are the 
holographic charges.

The paper is organized as follows. In the next section 
we review the definition of asymptotically locally AdS spacetimes.
In Section 3, after reviewing the asymptotic analysis, 
we formulate the variational problem with Dirichlet boundary conditions
for AdS gravity and we demonstrate the need for the counterterms.  
In Section 4 we derive an alternative expression for the holographic
charges using Noether's theorem and we show that these are reproduced 
by the covariant phase space method of Wald et al. \cite{Lee&Wald,Wald&Zoupas}.
We then use these results in Section 5 to prove the first law
of black hole mechanics for any AlAdS spacetime. We conclude with
two examples in Section 6, namely the four dimensional Kerr-Newman-AdS
black hole and the five dimensional Kerr-AdS black hole, which provides 
an illustration of the role of the conformal anomaly. Various technical
results are collected in the appendices. In appendix \ref{Weyl-tensor}
we comment on the connection between the `conformal mass' of
Ashtekar and Magnon \cite{AM} and the holographic mass.

\section{Asymptotically locally AdS spacetimes}
\setcounter{equation}{0}

We briefly discuss in this section the definition of asymptotically 
locally AdS spacetimes. For more details we refer to 
\cite{Skenderis:2002wp}\footnote{Note that the asymptotically locally 
AdS spacetimes here were called asymptotically AdS spacetimes in 
\cite{Skenderis:2002wp}.} (see also the math reviews \cite{Graham, anderson}).
 
Recall that $AdS_{d+1}$ is the maximally symmetric 
solution of Einstein's equations with negative cosmological constant,
$\L=-d(d-1)/2 l^2$, where $l$ is the radius of $AdS_{d+1}$ 
(we set $l=1$ from now on; one can easily reinstate this factor in all 
equations by dimensional analysis). Its curvature tensor is given by
\be \label{curAdS}
R_{\m \n \k \l} = 
g_{\m \l} g_{\n \k} - g_{\k \m} g_{\n \l},
\ee
and it has a conformal boundary with topology $\mathbf{R}\times 
\mathbf{S}^{d-1}$, where 
$\mathbf{R}$ corresponds to the time direction\footnote{Strictly speaking, the 
time coordinate is compact but one can go to the universal cover $CAdS$ where 
the time coordinate is unfolded. In this paper the time coordinate 
is always considered non-compact (except when we discuss the Euclidean
continuation).}.
Asymptotically locally AdS (AlAdS) spacetimes are solutions 
of Einstein's equations with Riemann tensor approaching (\ref{curAdS})
asymptotically (in a sense to be specified shortly).
A more restrictive class of spacetimes are the ones which asymptotically 
become exactly AdS spacetimes; 
these were called Asymptotically AdS (AAdS) in \cite{AM,Henneaux:1984xu}.
One can easily specialize our results to AAdS spacetimes and we shall do so 
in order to compare with existing literature.

The spacetimes $\mathcal{M}$ we consider are in particular {\it 
conformally compact manifolds} \cite{penrose}. This means that 
$\mathcal{M}$ is the interior of a manifold-with-boundary 
$\Bar{\mathcal{M}}$,  and the  metric $g$ has a second order pole at the 
boundary $B=\pa \mathcal{M}$, but there exists a 
defining function (i.e. $z(B)=0$, $dz(B) \neq 0$ and $z(\mathcal{M})>0$)
such that\footnote{Note that in 
\cite{Skenderis:2002wp} the symbol $g$ was used for the unphysical metric,
but here we use instead  $\mathtt{g}$ and use $g$ to denote the bulk metric.} 
\be \label{gdef}
\mathtt{g} = z^2 g
\ee
smoothly extends to $\Bar{\mathcal{M}}$, $\mathtt{g}|_B = \mathtt{g}_{(0)}$, 
and is non-degenerate. A standard argument (see for instance
\cite{Witten:1998qj}) implies that the boundary $B$ is 
equipped with a conformal class of metrics 
and $\mathtt{g}_{(0)}$ is a representative of the conformal
class. A conformally compact manifold that is also Einstein (i.e. 
solves Einstein's equations) is by definition an asymptotically locally 
AdS spacetime.

The most general asymptotics of such spacetimes was  determined 
in \cite{FG} for pure gravity and their analysis extends 
straightforwardly to include matter with soft enough behavior
at infinity, see for instance \cite{dHSS,Taylor-Robinson:2000xw,holren,PS1}. 
Near the boundary,
one can always choose coordinates in which the metric 
takes the form\footnote{In most examples in the literature 
the odd coefficients $\mathtt{g}_{(2k+1)}$ vanish (except when 
$2 k +1 =d$, the boundary dimension). In such cases, it is more convenient 
\cite{HS} to use instead of $z$ a new radial coordinate $\rho=z^2$.},
\bea \label{coord}
&&ds^2=g_{\m \n} dx^\m dx^\n = \frac{d z^2}{z^2} + 
\frac{1}{z^2} \mathtt{g}_{ij}(x,z) dx^i dx^j, \nonu
&&\mathtt{g}(x,z)=\mathtt{g}_{(0)} + z \mathtt{g}_{(1)} \cdots + z^{d} 
\mathtt{g}_{(d)} + \mathtt{h}_{(d)} z^{d} \log z^2 + ... 
\eea
In these coordinates the conformal boundary is located at $z=0$
and $\mathtt{g}_{(0)}$ is a representative of the conformal structure.
The asymptotic analysis reveals that all coefficients shown above 
except the traceless and divergenceless part of $\mathtt{g}_{(d)}$
are locally determined in terms of boundary data. The logarithmic term
appears only in even (boundary) dimensions (for pure gravity; if matter fields
are included, then a logarithmic term can appear in odd dimensions as 
well \cite{dHSS}) and is proportional \cite{dHSS} to the metric variation of 
the integrated holographic conformal anomaly \cite{HS}. 
A simple computation shows that the Riemann tensor of (\ref{coord}) is of the 
form (\ref{curAdS}) up to a correction of order $z$. This continues to be true 
in the presence of matter if the dominant contribution
of their stress energy tensor as we approach the boundary comes from the 
cosmological constant.  This is true for matter that 
corresponds to marginal or relevant operators of the dual 
theory in the AdS/CFT 
duality.

A very useful reformulation of the asymptotic analysis can be 
achieved by observing that for AlAdS spacetimes the radial 
derivative is to leading order equal to the dilatation 
operator \cite{PS1,PS2}. 
Let us first write the metric (\ref{coord}) in the form
\be
ds^2 = dr^2 + \gamma_{ij}(x,r) dx^i dx^j,
\ee
where $z=\exp(-r)$. The statement is that 
\be
\partial_r = \delta_D + \co(e^{-r}),
\ee
with
\be
\d_D = \int d^d x 
\left(2 \g_{ij} \frac{\d}{\d \g_{ij}} + (\D_I-d) \F^I \frac{\d}{\d \F^I} 
+ \cdots \right),
\ee
where we have included the contribution of massive scalars $\F^I$ dual to
operators of dimension $\D_I$ and the dots indicate the contribution
of other matter fields. The asymptotic analysis can now be very effectively
performed \cite{PS1} by expanding all objects in eigenfunctions of the 
dilatation operator and organizing the terms in the field equations according
to their dilatation weight.
 
Let us now see how $AdS_{d+1}$ and $AAdS_{d+1}$ spacetimes fit in this
framework. $AdS_{d+1}$ is conformally flat and this implies 
\cite{Skenderis:1999nb} that $\mathtt{g}_{(0)}$ is conformally flat as well and
the expansion (\ref{coord}) terminates at order $z^4$,
\be \label{AdScoef}
\mathtt{g}_{(4)} = \frac{1}{4} (\mathtt{g}_{(2)})^2, \qquad 
\mathtt{g}_{(2)ij} = -\frac{1}{d-2} (R_{ij} - \frac{1}{2 (d-1)} R 
\mathtt{g}_{(0)ij}),   
\ee
where $R_{ij}$ is the Ricci tensor of $\mathtt{g}_{(0)}$ 
($d=2$ is a special  case, see \cite{Skenderis:1999nb} for the 
expression of $\mathtt{g}_{(2)}$)
and $\mathtt{g}_{(0)}$ may be chosen to 
be the standard metric on $\mathbf{R} \times \mathbf{S}^{d-1}$.
Any $AAdS_{d+1}$ spacetime has a 
conformally flat representative $\mathtt{g}_{(0)}$ as well, which implies that 
all coefficients up to $\mathtt{g}_{(d)}$ 
are the same\footnote{In the presence of matter however 
these coefficients could acquire matter field dependence, 
see \cite{holren} for an example.} 
as for $AdS_{d+1}$, but $\mathtt{g}_{(d)}$ is different. In both cases, 
the logarithmic term is absent. 

AlAdS spacetimes have an arbitrary conformal structure $[\mathtt{g}_{(0)}]$ 
and a general $\mathtt{g}_{(d)}$, the logarithmic term is in general 
non-vanishing, and there is no 
{\it a priori} restriction on the topology of the conformal boundary. 
The mathematical structure of these spacetimes (or their Euclidean
counterparts) is under current investigation in the mathematics
community, see \cite{anderson}
and references therein. For instance, it is has not yet been established 
how many, if any, global solutions exist given a conformal structure, 
although given (sufficiently
regular) $\mathtt{g}_{(0)}$ and $\mathtt{g}_{(d)}$ a unique solution exists 
in a thickening
$B \times [0,\epsilon)$ of the boundary $B$. 
On the other hand, 
interesting examples of such spacetimes exist,
see \cite{anderson} for a collection of examples.

There is an important difference between even and odd dimensions.
When the spacetime is odd dimensional, there is a conformally 
invariant quantity $A[\mathtt{g}_{(0)}]$ one can construct 
using the boundary conformal structure $[\mathtt{g}_{(0)}]$, 
namely the integral of the holographic conformal anomaly \cite{HS}
(called renormalized volume in the math literature \cite{Graham})\footnote{
When certain matter fields are present one has additional conformal 
invariants in all dimensions, namely the matter conformal 
anomalies \cite{PS}; these play
a similar role to the gravitational conformal anomaly.}.
The conformal anomaly was found in \cite{HS} by considering the 
response of the 
{\em renormalized} on-shell action to Weyl transformations:
in order to render finite the on-shell gravitational action (which diverges
due to the infinite volume of the AlAdS spacetime) one is forced to 
add a certain number of boundary counterterms and the latter induce
an anomalous Weyl transformation. We will soon rediscover the need for 
counterterms and the anomaly via a different argument. 
For now we note that classically, the bulk metric 
determines a conformal  structure $[\mathtt{g}_{(0)}]$, and in odd dimensions, 
the latter determines a conformal invariant, the integrated anomaly 
$A[\mathtt{g}_{(0)}]$.

In the next section we investigate the variational problem
for AlAdS spacetimes. Given that a bulk metric is associated
with a conformal structure at infinity, we would like to formulate 
the variational problem such that the conformal structure is 
kept fixed. We will show that this is indeed possible when the 
anomaly vanishes.
When the anomaly is non-vanishing, however, the variational 
problem is more subtle: instead of keeping fixed a conformal
class, one chooses a representative and arranges such that 
the dependence of the theory on different representatives
is determined by the conformal class via the conformal anomaly.
In other words, we need to choose a representative in order to define 
the theory but the difference between different choices is 
governed by the conformal class. We will see that in all cases
the variational problem requires new boundary terms and these
are precisely the counterterms!

\section{Counterterms and the variational problem for AdS gravity} 
\label{variational}
\setcounter{equation}{0}

\subsection{The theory}

We will consider in this section the variational problem for 
AdS gravity coupled to scalars and a Maxwell field. Other matter 
fields, like forms and non-abelian gauge fields, can be easily incorporated
in the analysis, but for simplicity we do not include them. Moreover, to
keep the analysis general we do not include any Chern-Simons terms since
their particular form depends on the spacetime dimension. 
Within this framework we consider the most general Lagrangian 
consistent with the fact that the field equations admit 
a solution that is asymptotically locally AdS. 
The Lagrangian $D$-form ($D{=}d+1$) is given by
\be\label{Lagrangian}
{\bf L}=\left(\frac{1}{2\k^2}R-V(\F)\right)\ast{\bf 1}-
\frac12G_{IJ}(\F){\rm d}\F^I\wedge\ast {\rm d}\F^J-
\frac12 U(\F){\bf F}\wedge\ast {\bf F},
\ee   
where we use mostly plus signature and
${\bf F}={\rm d}{\bf A}$ and $V(\F),\, U(\F)$  and
$G_{IJ}(\F)$ are only constrained by the requirement that
the field equations admit AlAdS solutions. The exact conditions follow
from the asymptotic analysis discussed in the next subsection,
but we will not need the detailed form of the conditions
in this paper.
 
The variation of the
Lagrangian with respect to arbitrary field variations takes the form
\be\label{variation}
\d{\bf L}={\bf E}\d\psi+{\rm d}{\bf \Theta}(\psi,\d\psi),
\ee
where we use $\psi=(g_{\m\n},A_\m,\F^I)$ to denote collectively all
fields and ${\bf E}$ is the equations of motion $D$-form. More specifically,
we have
\be
\d{\bf L}={\bf E}^{\m\n}\sub{1}\d g_{\m\n}+{\bf E}^{\m}\sub{2}\d A_{\m}+
{\bf E}_{I}^{(3)}\d \F^{I}+{\rm d}{\bf \Theta}(\psi,\d\psi),
\ee  
where
\bea
&&{\bf E}\sub{1}^{\m\n}=-\frac{1}{2\k^2}\left(R^{\m\n}-\frac12 R g^{\m\n}-
\k^2\tilde{T}^{\m\n}\right)\ast{\bf 1}, \nonumber \\
&&{\bf E}\sub{2}^\n=\nabla_\m(U(\F)F^{\m\n})\ast{\bf 1}, \\
&&{\bf E}^{(3)}_I=\left(\nabla^\m(G_{IJ}(\F)\pa_\m\F^J)-\frac12
\frac{\pa G_{JK}}{\pa\F^I}
\pa_\m\F^J\pa^\m\F^K-\frac{\pa V}{\pa\F^I}-\frac14\frac{\pa U}{\pa \F^I}
F_{\m\n}F^{\m\n}\right)\ast{\bf 1}, \nonumber
\eea
and the matter stress tensor is given by
\bea\label{stresstensor}
\tilde{T}_{\m\n}=G_{IJ}(\F)\pa_\m\F^I\pa_\n\F^J+U(\F)F_{\m\r}F_\n\phantom{}^\r
-g_{\m\n}\mathcal{L}_{\rm m},
\eea
with $\mathcal{L}_{\rm m}$ denoting the matter part of the Lagrangian.
Moreover, 
\be\label{Theta}
{\bf \Theta}(\psi,\d\psi)=-\ast {\bf\it v}(\psi,\d\psi),
\ee
where
\be \label{v}
v^\m=-\frac{1}{2\k^2}(g^{\m\r}\nabla^\s\d g_{\r\s}-
g^{\r\s}\nabla^\m\d g_{\r\s})+G_{IJ}(\F)\d\F^I\nabla^\m\F^J+
U(\F)F^{\m\n}\d A_\n.
\ee

\subsection{Asymptotic analysis}

In this section we discuss the asymptotic solutions to the field 
equations. To formulate the problem we use a radial coordinate $r$ emanating 
orthogonally from the boundary in order to foliate spacetime into 
timelike hypersurfaces $\S_r$ diffeomorphic to the conformal boundary 
$\pa\mathcal{M}$. This can always be done at least in the vicinity of
the boundary. We then regulate the theory by introducing a cut-off 
hypersurface $\S_{r_o}$.

The most convenient way to perform the asymptotic analysis is by using 
a `radial Hamiltonian analysis' where the radial coordinate plays 
the role of time \cite{PS1} 
(see \cite{Kraus:1999di,deBoer:1999xf,Martelli:2002sp}
for earlier work). In this formalism, one uses the Gauss-Codazzi equations
to express the bulk equations of motion in terms of quantities intrinsic to
the radial hypersurfaces $\S_r$. In  the gauge
\beq\label{gauge}
ds^2=dr^2+\g_{ij}(r,x)dx^idx^j,\,\,\,A_r=0,
\eeq
the resulting equations of motion are given in appendix \ref{gauge_fixing}
and can be viewed as Hamilton's equations for the `radial canonical momenta',
\be \label{momdef}
\pi^{ij} =  -\frac{1}{2 \k^2}\sqrt{-\g} (K^{ij}-K\g^{ij}), \qquad 
\p^i =-\sqrt{-\g}  U(\F) \dot{A}^{i}, \qquad
\p^I = -\sqrt{-\g} G_{IJ}(\F)\dot{\F}^I, 
\ee 
where $K_{ij}=\half \dot{\g}_{ij}$ is the second fundamental form of the 
hypersurfaces $\S_r$
and the dot denotes differentiation w.r.t. the radial coordinate.

Within this framework, one is also able to express the on-shell value of the 
regulated action as an integral over the surface $\S_{r_o}$ by introducing a 
$\S_r$-covariant variable $\l$ such that
\be\label{lambda}
\int_{\mathcal{M}_{r_o}}{\bf L}_{\rm on-shell}=\int_{\mathcal{M}_{r_o}}d^{d+1}x
\sqrt{-g}\left(
\mathcal{L}_{\rm m}-\frac{1}{d-1}\tilde{T}^\s_\s\right)\equiv-\frac{1}{\k^2}
\int_{\S_{r_o}}d^dx\sqrt{-\g}\l.
\ee
Since $\S_{r_o}$ is compact, $\l$ is only defined up to a total divergence. 
We will soon fix this ambiguity by making a choice that 
simplifies the analysis. Taking the radial derivative of both sides of
(\ref{lambda}) we deduce that $\l$ must satisfy the differential equation 
\be\label{lambda_eq} 
\dot{\l}+K\l+\k^2\left(\mathcal{L}_{\rm m}-\frac{1}{d-1}
\tilde{T}^\s_\s\right)=0.
\ee
The regulated on-shell action (with the Gibbons-Hawking term included) is
then given by
\be\label{regac}
I_{r_o} = \frac{1}{\k^2} \int_{\S_{r_o}} d^d x \sqrt{-\g} (K-\l).
\ee
The radial momenta are now related to the on-shell action 
(see for instance \cite{PS1} and (\ref{pullback_theta}) below) by
\be \label{momenta}
\p^{ij}=\frac{\d I_{r_o}}{\d\g_{ij}},\qquad \p^{i}=\frac{\d I_{r_o}}{\d A_i},
\qquad \p_I=\frac{\d I_{r_o}}{\d\F^I}.
\ee
These expressions can be utilized to fix the total divergence ambiguity in
$\l$. In particular, since $I_{r_o}$ is unaffected by the addition of a
total divergence to $\l$, so are the momenta. We now argue that by adding 
an appropriate total derivative term to $\l$ we can always ensure that the
identity
\be\label{identity}
\p^{ij}\d\g_{ij}+\p^i\d A_i+ \p_I\d\Phi^I=\frac{1}{\k^2}\d\left[
\sqrt{-\g} (K-\l)\right]
\ee
holds {\em without} the integral over $\S_{r_o}$. This can always be achieved 
by the following procedure\footnote{\label{ft_var} 
This argument holds only for the 
{\em local} part of $\l$ and not for the non-local part $\l\sub{d}$ 
(see (\ref{momentum_exp}) below for the definition of this term) which
satisfies only the integrated version of (\ref{identity}). However, for
the special case of dilatations, $\d=\d_D$, $\l\sub{d}$ does satisfy 
(\ref{identity}). To see 
this consider an infinitesimal Weyl transformation of the renormalized action
(\ref{ren_action}): $\d_\s I_{\rm ren}=-\frac{2}{\k^2}\int_{\pa\cm}\sqrt{-\g}
(\tilde{K}\sub{d}-\tilde{\l}\sub{d})\d\s$. But from the renormalized version 
of (\ref{momenta}) we also have: $\d_\s I_{\rm ren}=\int_{\pa\cm}d^dx\sqrt{-\g}
\left[2\pi\sub{d}^i_i+(\D_I-d)\p\sub{\D_I}_I\F^I\right]\d\s$. Since $\d\s$ is
arbitrary, we can equate the integrands, which gives the same result as
(\ref{identity}) when specialized to $\d_D$.  }. 
Take first any $\l$ satisfying the definition (\ref{lambda}). The variation
$\d\left[\sqrt{-\g} (K-\l)\right]$ will then
generically produce terms with derivatives acting on the variations of the 
induced fields $\d\g_{ij}$, $\d A_i$ and $\d\Phi^I$. These derivatives can 
be moved to the coefficients of the field variations by integration by parts.
When all derivatives acting on the field variations are removed, 
(\ref{momenta}) guarantees that the coefficients of the field variations are
precisely the radial momenta. Now, the total derivative terms which are 
produced by this procedure can be absorbed into $\l$. In writing 
(\ref{identity}), we assume that such a procedure has been performed.

To carry out the asymptotic analysis we use the fact that asymptotically 
the radial derivative becomes equal to the dilatation operator, which 
with the current field content takes the form
\be
\d_D = \int d^d x 
\left(2 \g_{ij} \frac{\d}{\d \g_{ij}} 
+ \sum_I (\D_I-d) \F^I \frac{\d}{\d \F^I}\right).
\ee
This follows from the fact that on-shell, one can identify the radial 
derivative with the functional differential operator
\be\label{radial_derivative}
\pa_r=\int d^dx\left(2K_{ij}[\g,A,\Phi]\frac{\d}{\d\g_{ij}}+
\dot{A}_i[\g,A,\Phi]\frac{\d}{\d A_i}+\dot{\Phi}^I[\g,A,\Phi]\frac{\d}
{\d\Phi^I}\right)
=\d_D+\co(e^{-r}),
\ee
where the asymptotic behavior of the fields has been used. This observation
motivates an expansion of the momenta and the on-shell action in eigenfunctions
of $\d_D$:
\beqa\label{momentum_exp}
\p^i_j&=&\sqrt{-\g}\left(\p\sub{0}^i_j+\p\sub{2}^i_j+\cdots+\p\sub{d}^i_j+
\tilde{\p}\sub{d}^i_j\log{\rm e}^{-2r}+\cdots\right), \nonumber \\
\p^i&=&\sqrt{-\g}\left(\p\sub{3}^i+\p\sub{4}^i+\cdots+\p\sub{d}^i+
\tilde{\p}\sub{d}^i\log{\rm e}^{-2r}+\cdots\right),\\
\p_I&=&\sqrt{-\g}(\sum_{d-\D_I\leq s< \D_I} \p\sub{s}_I+\p\sub{\D_I}+
\tilde{\p}\sub{\D_I}_I\log{\rm e}^{-2r}+\cdots), \NO \\
\l&=&\l\sub{0}+\l\sub{2}+\cdots+\l\sub{d}+
\tilde{\l}_{(d)}\log{\rm e}^{-2r}+\cdots. \nonumber 
\eeqa
All terms in these expansions transform under dilatations according to 
their subscript, e.g. $\p\sub{n}$ transforms as
\be
\d_D\p\sub{n} = - n \p\sub{n},
\ee
except for the normalizable (``vev'') part of the expansions, $\p\sub{d}^i_j$,
$\p\sub{d}^i,\, \p\sub{\D_I},\, \l\sub{d}$, which  transform inhomogeneously,
with the inhomogeneous term equal to ({-}2) the coefficient of the logarithmic
piece. For example,
\be
\d_D \p\sub{d}^i_j =  - d\p\sub{d}^i_j  - 2 \tilde{\p}\sub{d}^i_j. 
\ee 
(Note that the transformation of the volume element, 
namely $\d_D\sqrt{-\g}=d \sqrt{-\g}$, should be taken into account).

The significance of writing the radial derivative in the form 
(\ref{radial_derivative}) is that the above expansions imply that the
radial derivative can also be expanded in a series of covariant
functional operators, $\d\sub{n}$, of successively higher dilatation weight 
that commute with the dilatation operator. Namely,
\be\label{radial_exp}
\pa_r=\d_D+\d\sub{1}+\ldots
\ee
This allows us to perform the asymptotic analysis in a covariant way by 
substituting the expansions (\ref{momentum_exp}) and the expansion for
the radial derivative in the field equations and collecting terms with the 
same weight. This determines uniquely and {\em locally} all coefficients of 
momenta in (\ref{momentum_exp}), except for the traceless and divergenceless 
part of $\p\sub{d}^i_j$ and the divergenceless part of $\p\sub{d}^i$, in terms 
of the boundary data, i.e. in terms of the  induced fields on $\S_{r_o}$. 
\begin{flushleft}
{\em Ward identities}
\end{flushleft}

The divergence of $\p\sub{d}^i_j$ and $\p\sub{d}^i$, which are determined
respectively by the second equation in (\ref{gf_einstein}) and the first 
equation in  (\ref{gf_vector}), yield the Ward identities related to
boundary diffeomorphisms and $U(1)$ gauge transformations respectively.
These read,
\bea \label{WI}
&&2D_i\pi\sub{d}^i_j +\p\sub{d}^iF_{ij}-\pi_I\pa_j\Phi^I=0, \NO\\
&&D_i \p\sub{d}^i =0. 
\eea

The trace Ward identity follows from the explicit expression of 
$\p\sub{d}^i_i$ obtained by solving asymptotically the 
field equations. Alternatively, one can follow the argument in footnote 
\ref{ft_var}. As explained, the result is identical to that obtained by
 specializing (\ref{identity}) to dilatations and considering the terms of 
weight $d$ - although only the integrated version of (\ref{identity}) 
holds for these terms. The result is 
\be\label{traceWI}
2\p\sub{d}^i_i+\sum_I(\D_I-d)\p\sub{\D_I}_I\Phi^I=
-\frac{2}{\k^2} (\tilde{K}\sub{d}-\tilde{\l}\sub{d})\equiv \ca,
\ee 
where $\ca$ is the trace anomaly  \cite{HS}.
\begin{flushleft}
{\em Renormalized action}
\end{flushleft}

Finally, the renormalized action is defined as\footnote{We will often refer
to the same quantity evaluated on $\S_{r_o}$  as the `renormalized action', 
i.e. before the limit $r_o\to\infty$ is taken.} 
\be\label{ren_action}
I_{{\rm ren}} = \lim_{r_o \to \infty} (I_{r_o} + I_{{\rm ct}}) = 
\frac{1}{\k^2}\int_{\partial \mathcal{M}} d^d x \sqrt{-\g} (K\sub{d} - 
\l\sub{d}),
\ee
where the counterterm action, $I_{{\rm ct}}$, is given by
\be \label{count}
I_{{\rm ct}} = -\frac{1}{\k^2} \int_{\S_{r_o}} d^d x \sqrt{-\g} \left( 
\sum_{n=0}^{d-1} (K_{(n)}-\l_{(n)})+(\tilde{K}_{(d)} - \tilde{\l}_{(d)})
\log e^{-2 r_o} \right).
\ee

\subsection{Gauge invariance of the renormalized action}
\label{gauge_invariance}

In this section we first determine the most general bulk diffeomorphisms and 
$U(1)$ gauge transformations which preserve the gauge  
(\ref{gauge}). We note that this gauge need only  be preserved
up to terms of next-to-normalizable mode order, i.e. up to order 
$e^{-(d-1)r}$. Such transformations leave invariant the functional form
of the boundary conditions, of the asymptotic solutions,
and of the counterterm action on
the regulated boundary $\S_{r_o}$. Subsequently, we derive the maximal 
subset of  gauge-preserving transformations that leave the renormalized 
action invariant, where only the functional form of the boundary 
conditions is imposed, namely we require that 
\be \label{leading}
\g_{ij}(x,r) \sim e^{2 r} \mathtt{g}_{(0)ij}(x), 
\qquad A_i (x,r) \sim \mathtt{A}_{(0)i}(x), \qquad
\F^I (x,r) \sim \mathtt{\f}^I_{(0)}(x) e^{-(d-\D_I)r},
\ee
but no conditions are imposed on $\mathtt{g}_{(0)ij},\; \mathtt{A}_{(0)i},\; 
\mathtt{\f}^I_{(0)}$. Notice that the transformations below do 
act on these coefficients.

In the gauge (\ref{gauge}), the Lie derivative, 
$\mathcal{L}_\xi$, of the bulk fields w.r.t. a bulk vector field $\xi^\m$ is 
given by   
\bea
&&\mathcal{L}_\xi g_{rr}=\dot{\xi}^r,\NO\\
&&\mathcal{L}_\xi g_{ri}=\g_{ij}(\dot{\xi}^j+\pa^j\xi^r),\NO\\
&&\mathcal{L}_\xi g_{ij}=L_\xi\g_{ij}+2K_{ij}\xi^r\sim L_\xi\g_{ij}+2\g_{ij}
\xi^r,\label{lg} \\\NO\\
&&\mathcal{L}_\xi A_r=A_j\dot{\xi}^j,\NO\\
&&\mathcal{L}_\xi A_i=L_\xi A_i+\xi^r \dot{A}_i\sim L_\xi A_i,
\label{la} \\\NO\\
&&\mathcal{L}_\xi\Phi^I=L_\xi\Phi^I+\xi^r\dot{\F}^I\sim L_\xi\Phi^I+(\D_I-d)
\xi^r\F^I, \label{lf}
\eea
where $L_\xi$ is the Lie derivative w.r.t. the transverse components $\xi^i$
of the bulk vector field $\xi$. This bulk diffeomorphism, combined with a 
$U(1)$ gauge transformation, preserves the gauge fixing (up to the desired 
order; see (\ref{order}) below) provided 
$\mathcal{L}_\xi g_{rr}=\mathcal{L}_\xi g_{ri}=\co(e^{-dr})$ and
$\mathcal{L}_\xi A_r+\dot{\a}=\co(e^{-(d+2)r})$. Integrating these conditions 
gives:
\bea\label{PBH_transformation}
&&\xi^r=\d\s(x)+\co(e^{-dr}),\NO\\
&&\xi^i=\xi_o^i(x)+\pa_j\d\s(x)\int_r^\infty dr'\g^{ji}(r',x)+\co(e^{-(d+2)r}),
\NO\\
&&\a=\a_o(x)+\pa_i\d\s(x)\int_r^\infty dr'A^i(r',x)+\co(e^{-(d+2)r}),
\eea
where $\d\s(x)$ and $\a_o(x)$ are arbitrary functions of the transverse
coordinates and $\xi_o^i(x)$ is an arbitrary transverse vector field.
For $\xi_o=0$, this bulk diffeomorphism is precisely the 
`Penrose-Brown-Henneaux (PBH) transformation'
\cite{penrose, Brown:1986nw}
which induces a Weyl transformation on the conformal boundary
\cite{Imbimbo:1999bj, dHSS, Skenderis_proceedings}. Here,
we will call a `PBH transformation' 
the combined bulk diffeomorphism with
$\xi_o=0$ {\em and} the gauge transformation with $\a_o=0$, which is
required in order to preserve the gauge of the Maxwell field.

Next we  determine which subset of (\ref{PBH_transformation}) leaves
invariant the renormalized action 
\be
I_{\rm ren}=\int_{\cm_{r_o}}{\bf L}
+\frac{1}{\k^2}\int_{\S_{r_o}}d^dx\sqrt{-\g}K
+I_{\rm ct},
\ee
where 
\be
{\bf L}=\left(\frac{1}{2\k^2}R[g]+\cl_{\rm m}\right)\ast{\bf 1},
\ee 
and $I_{\rm ct}$ is given by (\ref{count}). Since ${\bf L}$ is 
covariant under diffeomorphisms and gauge invariant, we have
\be
\d_\xi{\bf L}=\cl_\xi{\bf L}={\rm d} i_\xi{\bf L},\,\,\,\, 
\d_\a{\bf L}=0,
\ee
where we have used the identity $\cl_\xi=i_\xi{\rm d}+{\rm d}i_\xi$ for the 
Lie derivative on forms. Hence,
\be
\d_{\xi,\a} I_{\rm ren}=\int_{\S_{r_o}}d^dx\sqrt{-\g}\xi^r\left(\frac{1}{2
\k^2}R[g]+\cl_{\rm m}\right)+\frac{1}{\k^2}\d_\xi\int_{\S_{r_o}}d^dx\sqrt{-\g}
K+\d_{\xi,\a} I_{\rm ct}.
\ee  
Now, in the gauge we are using, the  Ricci scalar of the bulk metric can 
be expressed as
\be
R[g]=R+K^2-K_{ij}K^{ij}-\frac{2}{\sqrt{-\g}}\pa_r(\sqrt{-\g}K),
\ee
Moreover, for the diffeomorphisms given by (\ref{PBH_transformation}) a short
computation gives
\be
\d_\xi\int_{\S_{r_o}}d^dx\sqrt{-\g}K=\int_{\S_{r_o}}d^dx\xi^r\pa_r(\sqrt{-\g}
K),
\ee
and hence
\be
\d_{\xi,\a} I_{\rm ren}=\frac{1}{2\k^2}\int_{\S_{r_o}}d^dx\sqrt{-\g}\xi^r
\left(R+K^2-K_{ij}K^{ij}+2\k^2\cl_{\rm m}\right)+\d_{\xi,\a} I_{\rm ct}.
\ee   
The last term takes the form
\be
\d_{\xi,\a} I_{\rm ct}=\int_{\S_{r_o}}d^dx\left(\hat{\p}_{\rm ct}^{ij}
\d_\xi\g_{ij}
+\hat{\p}_{\rm ct\, I}\d_\xi\F^I+\hat{\p}_{\rm ct}^i(\d_\xi A_i+\pa_i\a)
\right),
\ee
where we put hats on the counterterm 
momenta to emphasize that they should be viewed as predetermined local 
functionals of the induced fields as opposed to the asymptotic behavior of 
the radial derivative of the on-shell induced fields.
Inserting now the transformation (\ref{PBH_transformation}) and using 
the second equation in (\ref{gf_einstein}) and the first equation in
(\ref{gf_vector}), which the counterterms satisfy by construction, we
are left with
\bea\label{gauge_variation}
\d_{\xi,\a} I_{\rm ren}=\int_{\S_{r_o}}d^dx\xi^r\left\{
\frac{1}{2\k^2}\sqrt{-\g}\left(R+K^2-K_{ij}K^{ij}+2\k^2\cl_{\rm m}\right)
\phantom{more}\right.\NO\\\left.+\left(\hat{\p}_{\rm ct}^{ij}2K_{ij}
+\hat{\p}_{\rm ct\, I}\dot{\F}^I+\hat{\p}_{\rm ct}^i\dot{A}_i\right)\right\}.
\eea
Using the form of the boundary conditions (\ref{leading}) one finds that 
the leading order divergent term cancels and the terms inside the
curly brackets are of order $e^{(d-1) r}$.
We therefore conclude that a transformation (\ref{PBH_transformation}) 
that leaves the renormalized action invariant must have 
\be\label{order}
\xi^r=\co(e^{-dr}),\qquad \dot{\xi}^i=-\pa^i\xi^r+\co(e^{-(d+2) r}) 
=\co(e^{-(d+2) r}).
\ee
This leaves us with $\xi^i=\xi_o^i(x)$ and
$\a=\a_o(x)$, up to sufficiently high order in $e^{-r}$ as $r\to\infty$.    

In fact, as is well known, the PBH transformation, i.e. the part of the 
transformation (\ref{PBH_transformation}) that is driven by $\d\s(x)$,  
induces a Weyl transformation on the boundary and even  the
{\em on-shell} renormalized action is not invariant under such 
transformations unless the anomaly vanishes. To see this
let us first rewrite the Hamilton constraint (first equation in 
(\ref{gf_einstein})) as
\be
\frac{1}{2\k^2}\sqrt{-\g}\left(R+
K^2-K_{ij}K^{ij}+2\k^2\cl_{\rm m}\right)=\p^{ij}2K_{ij}
+\p_I\dot{\F}^I+\p^i\dot{A}_i.
\ee
Then, using the trace Ward identity (\ref{traceWI}), 
(\ref{gauge_variation}) becomes on-shell 
\be
\d_{\xi,\a} I_{\rm ren}^{\rm on-shell}
=\int_{\S_{r_o}}d^dx\sqrt{-\g}\xi^r\ca.
\ee

\subsection{Variational problem}
\label{variational_problem}

We investigate in this section under which conditions
the variational problem is well-posed, i.e. under which 
conditions the boundary 
terms in the variation of the action cancel so that
$\d I = 0$ (under generic variation) 
implies the field equations and vice versa.

Let $n^\m$ be the outward unit normal to the hypersurfaces $\S_r$. 
Using (\ref{v}) and the definition of the radial momenta
(\ref{momdef}) one easily finds that the pullback of ${\bf \Theta}$ onto
$\S_r$ is given by\footnote{Up to an exact term ${\rm d}(\ast{\bf y})$, 
where $y^\m=\frac{1}{2\k^2}n^\r g^{\m\s}\d g_{\r\s}$ vanishes for variations 
that preserve the gauge fixing.} 
\bea\label{pullback_theta}
{\bf\Theta} & = & -n_\m v^\m \ast_\S{\bf 1}\NO\\
&=& \left(-\frac{1}{\k^2}\d(\sqrt{-\g}K)+
\p^{ij}\d\g_{ij}+\p^i\d A _i+\p_I\d\F^I\right)d\m,
\eea
where $\sqrt{-\g}d\m\equiv\ast_\S{\bf 1}$, and $\ast_\S$ denotes 
the Hodge dual 
w.r.t. $\S_r$. We thus arrive at the well-known fact \cite{GH} that the 
Gibbons-Hawking term is sufficient to render  the variational problem 
well-defined when all induced fields at the boundary are kept fixed, i.e.
\be \label{regbc}
\d\g_{ij}=0,\,\,\,\d A_i=0,\,\,\,\d\Phi^I=0 \,\,\,\,{\rm on}\,\, \S_{r_o}.
\ee
These boundary conditions  are perfectly acceptable 
for the regulated manifold with boundary $\S_{r_o}$ at finite $r_o$, since 
the bulk fields uniquely induce fields on $\S_{r_o}$.
However, as $\S_{r_o}\to\pa\mathcal{M}$ this is no longer the case.
The induced fields generically diverge (or vanish) in this limit and
the bulk fields only determine the {\em conformal class} of the
boundary fields. It is therefore not possible to impose the above 
boundary conditions on the conformal boundary. 
At most, one can demand that the boundary fields are kept fixed up to a 
 Weyl transformation, namely
\be\label{boundary_condition}
\d\g_{ij}=2\g_{ij}\d\s,\,\,\,\d A_i=0,\,\,\,\d\F^I=(\D_I-d)\F^I\d\s\,\,\,\,
{\rm on}\,\,\pa\cm.
\ee
To implement these weaker boundary conditions we 
insert the expansions (\ref{momentum_exp}) into (\ref{pullback_theta}) and 
use (\ref{identity}) to get
\bea\label{pullback_theta_exp}
{\bf\Theta}& = & \left\{-\frac{1}{\k^2}\d(\sqrt{-\g}K)
-(\p_{\rm ct}^{ij}\d\g_{ij}+\p_{\rm ct}^i\d A_i+\p_{{\rm ct}I}\d\F^I)
\right.\NO\\
&&\left.\phantom{more}+\sqrt{-\g}(\p\sub{d}^{ij}\d\g_{ij}+\p\sub{d}^i\d A_i+
\p\sub{\D_I}_I\d\F^I)+\ldots\right\}d\m\NO\\
& = &  \left\{\d\left(-\frac{1}{\k^2}\sqrt{-\g}[K-(K-\l)_{\rm ct}]\right)
\right.\NO\\
&&\left.\phantom{more}+\sqrt{-\g}(\p\sub{d}^{ij}\d\g_{ij}+\p\sub{d}^i\d A_i+
\p\sub{\D_I}_I\d\F^I)+\ldots\right\}d\m.
\eea
Hence,
\be
\int_{\S_{r_o}}{\bf\Theta} 
= \d \left(-\frac{1}{\k^2}\int_{\S_{r_o}}d^dx\sqrt{-\g} K{-}I_{\rm ct}\right)
{+}\int_{\S_{r_o}}d^dx\sqrt{-\g}\left[\pi\sub{d}^{ij} \d \g_{ij} 
+ \p\sub{d}^i \d A_i + \p\sub{\D_I}_I \d \F^I +
\ldots\right], \qquad
\ee
where the dots denote terms of higher dilatation weight which
do not survive after the regulator is removed and $I_{\rm ct}$ is {\em local} 
in the boundary fields. Finally  we insert the boundary conditions 
(\ref{boundary_condition}) and use the diffeomorphism and trace Ward 
identities (\ref{WI}) and (\ref{traceWI}) to arrive at
\be\label{anomaly}
\int_{\S_{r_o}}{\bf\Theta} 
= \d \left(- \frac{1}{\k^2}\int_{\S_{r_o}}d^dx\sqrt{-\g}K{-}I_{\rm ct}\right)
{+}\int_{\S_{r_o}}d^dx\sqrt{-\g}\mathcal{A}\d\s.
\ee
It follows that 
\be\label{anomaly_2}
\d I^{{\rm on-shell}}_{{\rm ren}} = 
\int_{\S_{r_o}}d^dx\sqrt{-\g}\mathcal{A}\d\s.
\ee
Notice that $\mathcal{A}$ is uniquely determined from boundary data.
Furthermore, its integral is conformally invariant. It follows that 
$\mathcal{A}$ is a conformal density of weight $d$ modulo total derivatives.

There are three cases to discuss now. 
\begin{enumerate}
\item The unintegrated anomaly vanishes identically: 
\be
\ca \equiv 0.
\ee
This is the case, for instance, for pure AAdS gravity in even dimensions.
Our analysis shows that the variational problem in this case
is well-posed, provided we augment the Gibbons-Hawking term by the 
usual counterterms. 
\item The integrated anomaly vanishes for a particular conformal class
$[\mathtt{g}\sub{0}]$,
\be\label{int_anomaly}
A[\mathtt{g}\sub{0}] \equiv \int_{\pa \mathcal{M}} d^dx
\sqrt{-\mathtt{g}\sub{0}} \ca[\mathtt{g}\sub{0}]=0.
\ee
This is the case, for instance, for pure AAdS gravity in odd dimensions 
with the conformal class represented by the standard metric on the 
boundary $\mathbf{R}\times\mathbf{S}^{d-1}$. 
When (\ref{int_anomaly}) holds the anomaly density does not necessarily vanish 
and so the variational problem with the boundary conditions 
(\ref{boundary_condition}) is 
not well-defined in general since the variation of the action generically 
contains a non-vanishing boundary term. Nevertheless, the vanishing of the 
integrated anomaly guarantees that there exists a representative 
$\mathtt{g}_{(0)}$ of the 
conformal class $[\mathtt{g}\sub{0}]$ 
for which the anomaly density, $\mathcal{A}$, is zero. 
For instance, for pure AAdS gravity in odd dimensions such a representative
is the standard metric on $\mathbf{R}\times\mathbf{S}^{d-1}$.
Hence one can pick a suitable defining function which induces this
particular representative.  In practice this means that we want to perform 
a PBH transformation such that the resulting radial coordinate acts as
a defining function which induces the desired representative. We then 
consider the variational problem {\em around this gauge} that
corresponds to the privileged representative of the conformal structure
at the boundary for which the anomaly density vanishes. However, this ensures
only that the first order variation of the action will contain no boundary 
terms. To make the variational problem well-defined to all orders one is forced
to break the bulk diffeomorphisms which induce a Weyl transformation on
the boundary and consider variations of the bulk fields which preserve
a particular representative of the conformal class. In other words, in
this case, in order to make the variational problem well-defined to
all orders we {\em must} impose the boundary conditions, 
\be \label{anbc}
\d \mathtt{g}_{(0)ij}=0, \qquad \d \mathtt{A}_{(0)i}=0, \qquad 
\d \phi_{(0)}=0, \qquad 
\ee
where $\mathtt{g}_{(0)ij}(x)$ is the chosen representative of the conformal
structure and 
$\mathtt{A}_{(0)i}(x)$ and $\mathtt{\phi}_{(0)}(x)$ are the leading 
terms in the asymptotic expansion of the bulk gauge and scalar 
fields, respectively,
\be 
A_i(x,r) = \mathtt{A}_{(0)i}(x)(1 + \co(e^{-r})), \qquad
\Phi(x,r) = \phi_{(0)}(x) e^{-(d-\D_I) r} (1 + \co(e^{-r})).
\ee
As we have seen, however, this is only possible if one breaks certain bulk
diffeomorphisms. 
\item The integrated anomaly is non-zero. In this case, to ensure that the
variational problem is well defined already at leading order, we have to pick 
a representative and allow only variations that preserve the corresponding  
gauge.

\end{enumerate}

To summarize, we have seen that bulk covariance in AlAdS spaces requires
that we formulate the variational problem with the boundary conditions
(\ref{boundary_condition}) instead of the stronger (\ref{regbc}). The
counterterms are {\em essential} in making the variational problem
well-defined  with such boundary conditions and are exactly on the same
footing with the Gibbons-Hawking term. However, when the unintegrated anomaly
does not vanish identically, the variational problem can only be well-defined
(to all orders) with the boundary conditions (\ref{anbc}), which can only
be imposed if certain bulk diffeomorphisms are broken. The counterterms in
this case guarantee that the on-shell action has a well-defined transformation
under the broken diffeomorphisms. The transformation is given precisely by
the anomaly.

\section{Holographic charges are Noether charges}
\setcounter{equation}{0}
\label{conserved_charges}

\subsection{Conserved charges associated with asymptotic symmetries}
\label{Noether_charges}
 
We have seen in section \ref{gauge_invariance} that the renormalized action
is invariant under bulk diffeomorphisms and $U(1)$ gauge transformations
that asymptotically take the form (\ref{PBH_transformation}) provided 
$\xi^r=\co(e^{-dr})$. Moreover, requiring that such transformations 
preserve the boundary 
conditions (\ref{boundary_condition}) constrains $\x^i$ to be an asymptotic
conformal Killing vector, i.e. to asymptotically 
approach a boundary conformal Killing vector (see appendix \ref{KV} for 
the precise definition). 
When the anomaly does not vanish, however,
we impose the boundary conditions (\ref{anbc}) which are only preserved
if $\xi^i$ is a boundary {\em Killing} vector (as opposed to asymptotic 
conformal Killing vector).     

We now apply Noether's theorem to extract the conserved currents and charges 
associated with these asymptotic symmetries. To this end we first consider the
following field variations: 
\be\label{Noether_variations}
\d_1\psi=f_1(r,x)\cl_\xi\psi,
\quad\d_2\psi=f_2(r,x)\d_\a\psi,
\ee
where $f_1(r,x),\,f_2(r,x)$ are arbitrary functions on $\cm$ which reduce to 
functions $\bar{f}_1(x)$ and $\bar{f}_2(x)$ respectively on $\pa\cm$, $\xi^i$ 
is an asymptotic conformal Killing vector of the induced fields on $\S_r$ and 
$\a$ is a gauge parameter which asymptotically tends to a constant. These,
transformations are not a symmetry of the renormalized action unless $f_1$ and
$f_2$ are constants, but they preserve the boundary conditions 
(\ref{boundary_condition}) for arbitrary $f_1,\,f_2$. Varying the renormalized
action, whose general variation is given by
\be 
\delta I_{{\rm ren}} = \int_{\cm_{r_o}} {\bf E}\d\psi 
+ \int_{\S_{r_o}} d^dx\sqrt{-\g}\left(\pi\sub{d}^{ij} \d \g_{ij} 
+ \p\sub{d}^i \d A_i + \p\sub{\D_I}_I \d \F^I \right),
\ee
with respect to such field variations we will now derive the conserved Noether 
charges.
 
\begin{flushleft}
{\em Electric charge}
\end{flushleft} 

Let us first consider the transformation $\d_2\psi$ and derive the 
corresponding conserved current. Since $\d_\a{\bf L}=0$, we have from 
(\ref{variation})
\be
{\bf E}\d_\a\psi=-{\rm d}{\bf \Theta}(\psi,\d_\a\psi).
\ee
Hence,
\be
\d_2 I_{\rm ren} =-\int_{\cm_{r_o}}f_2{\rm d}{\bf \Theta}(\psi,\d_\a\psi) 
+\int_{\S_{r_o}}d^dx\sqrt{-\g}f_2\p\sub{d}^i\pa_i\a,
\ee
But $\a$ is asymptotically constant and so the boundary term vanishes. Hence, 
on-shell the bulk integral on the RHS must vanish for arbitrary $f_2$, which 
leads to the conservation law for the $U(1)$ current
\be
{\bf J}_\a\equiv{\bf \Theta}(\psi,\d_\a\psi).
\ee
Since on-shell ${\bf J}_\a$ is closed, it is locally exact. In fact one 
easily finds 
\be
{\bf J}_\a={\rm d}{\bf Q}_\a,
\ee
where ${\bf Q}_\a=-\a\ast\cf$ and $\cf_{\m\n}=U(\F)F_{\m\n}$. 
Then, given a Cauchy surface $C$, the conserved Noether charge is given 
by\footnote{\label{orientation} Throughout this article we use the convention 
about the (relative) orientation 
$\e_{r t i_2 \ldots i_d} \equiv \e_{t i_2 \ldots i_d} = +1$.
 The minus sign in the definition of the electric charge is included 
to compensate for this
choice of orientation, which is opposite from the conventional one.} 
 
\be \label{elch}
Q = \int_C {\bf J}_\a=-\int_{\pa\cm\cap C}\ast\mathcal{F},
\ee
where we have assumed without loss of generality  that $\a\to 1$ on $\pa\cm$. 
One can check that this charge is conserved, i.e. independent of the 
Cauchy surface $C$, which follows immediately from the field equation
\be
{\rm d}\ast\mathcal{F}=0.
\ee  

\begin{flushleft}
{\em Charges associated with boundary conformal isometries}
\end{flushleft}

The same argument can be applied to derive the conserved currents and
Noether charges associated with asymptotic conformal isometries of the 
induced fields. Again from (\ref{variation}) we have
\be
{\bf E}\cl_\xi\psi={\rm d}\left(i_\xi{\bf L}-{\bf\Theta}(\psi,\cl_\xi\psi)
\right).
\ee
Hence, defining the current
\be
{\bf J}[\xi]\equiv{\bf \Theta}(\psi,\mathcal{L}_\xi\psi)-i_\xi{\bf L},
\ee
we get
\be\label{CKV_variation}  
\delta_1 I_{{\rm ren}} = -\int_{\cm_{r_o}} f_1{\rm d}{\bf J}[\xi] 
+ \int_{\S_{r_o}}d^dx \sqrt{-\g}f_1\left(\pi\sub{d}^{ij} L_\xi\g_{ij} 
+ \p\sub{d} L_\xi A_i+ \p\sub{\D_I}_I L_\xi \F^I\right).
\ee
Since $\xi^i$ is an asymptotically conformal Killing vector, it follows
that
\be 
\delta_1 I_{{\rm ren}} = -\int_{\cm_{r_o}} f_1{\rm d}{\bf J}[\xi] 
+ \int_{\S_{r_o}}d^dx \sqrt{-\g}f_1\left(2\pi\sub{d}^i_i 
+(\D_I-d) \p\sub{\D_I}_I\F^I\right)\frac1d D_i\xi^i.
\ee
Now, evaluating the LHS using (\ref{anomaly_2}) and the RHS using the trace 
Ward identity (\ref{traceWI}),
we deduce that on-shell the bulk integral vanishes, which leads to
the conservation law
\be\label{conservation_CKV}
{\rm d}{\bf J}[\xi]=0.
\ee
Hence, ${\bf J}[\xi]$ is locally exact, ${\bf J}[\xi]={\rm d}{\bf Q}[\xi]$,
and it is easily shown that
\be\label{Xi}
{\bf Q}[\xi]=-\frac{1}{\k^2}\ast{\bf \Xi}[\xi],
\ee
where the 2-form ${\bf \Xi}$ is given by
\be
\Xi_{\m\n}=\nabla_{[\m}\xi_{\n]}+\k^2 U(\F)F_{\m\n}A_\r \xi^\r.
\ee
However, ${\bf J}[\xi]$ is not the full Noether current in this case as
there is an extra contribution with support on $\S_{r_o}$. To derive the
correct form of the current we use (\ref{pullback_theta_exp}) to rewrite 
(\ref{CKV_variation}) as
\bea
\d_1 I_{\rm ren} & = &  \int_{\cm_{r_o}}{\rm d} f_1\wedge{\bf J}[\xi]
-\int_{\S_{r_o}}f_1{\bf J}[\xi]
+\int_{\S_{r_o}}d^dx \sqrt{-\g}f_1\left(\pi\sub{d}^{ij} L_\xi\g_{ij} 
+ \p\sub{d} L_\xi A_i+ \p\sub{\D_I}_I L_\xi \F^I\right)\NO\\
&=& \int_{\cm_{r_o}}{\rm d} f_1\wedge{\bf J}[\xi]
+\int_{\S_{r_o}}f_1i_\xi{\bf L}+\int_{\S_{r_o}}d^dx f_1\d_\xi\left(\frac{1}{
\k^2}\sqrt{-\g}[K-(K-\l)_{\rm ct}]\right).
\eea
Since $\xi$ is tangent to $\S_{r_o}$, the second term vanishes. To put the
last term in the desired form, we define the $d$-form  
\be\label{B}
{\bf B}\equiv -\frac{1}{\k^2}[K-(K-\l)_{\rm ct}]\ast_\S{\bf 1}
\ee  
on $\S_r$ which is covariant w.r.t. diffeomorphisms within $\S_r$. Using
the identity $L_\xi=\bar{\rm d}i_\xi+i_\xi \bar{\rm d}$ on forms, $\bar{d}$
being the exterior derivative on $\S_r$, we obtain
\bea
\d_1 I_{\rm ren} & = &  \int_{\cm_{r_o}}{\rm d} f_1\wedge{\bf J}[\xi]
-\int_{\S_{r_o}}f_1\d_\xi{\bf B}\NO\\
& = & \int_{\cm_{r_o}}{\rm d} f_1\wedge{\bf J}[\xi]
-\int_{\S_{r_o}}f_1\bar{\rm d}i_\xi{\bf B}\NO\\
& = & \int_{\cm_{r_o}}{\rm d} f_1\wedge{\bf J}[\xi]
+\int_{\S_{r_o}}\bar{\rm d}f_1\wedge i_\xi{\bf B}\NO\\
& = & \int_{\cm_{r_o}}{\rm d} f_1\wedge{\bf J}[\xi]
+\int_{\cm_{r_o}}\r(\S_{r_o})\wedge{\rm d}f_1\wedge i_\xi{\bf B},
\eea
where $\r(\S_r)$ is a one-form with delta function support on $\S_r$,
known as the Poincar\'e dual of $\S_r$ in $\cm$. Therefore, the full Noether 
current is
\be
\tilde{\bf J}[\xi]\equiv{\bf J}[\xi]-\r(\S_{r_o})\wedge i_\xi{\bf B}.
\ee 
Given a Cauchy surface $C$, we now define the Noether charge
\be\label{BCKV_Noether_charge}
Q[\xi]\equiv\int_{C}\tilde{\bf J}[\xi]=\int_{\pa\cm\cap C}\left({\bf Q}[\xi]
-i_\xi{\bf B}\right).
\ee 
If $C$ and $C'$ are two Cauchy surfaces whose intersection with $\pa\cm$ bounds
a domain $\D\subset \pa\cm$, then Stokes' theorem and the conservation law
(\ref{conservation_CKV}) imply
\bea
Q_C[\xi]-Q_{C'}[\xi] & = &\int_{\D\subset \pa\cm}\left({\bf J}[\xi]-{\rm d}
i_\xi{\bf B}\right)\NO\\
& = & \int_{\pa\cm}d^dx \sqrt{-\g}\left(\pi\sub{d}^{ij} L_\xi\g_{ij} 
+ \p\sub{d} L_\xi A_i+ \p\sub{\D_I}_I L_\xi \F^I\right)\NO\\
& = &\int_{\pa\cm}d^dx \sqrt{-\g}\left(\pi\sub{d}^i_i+(\D_I-d)\p\sub{\D_I}_I
\F^I\right)\frac1d D_i\xi^i\NO\\
& = &\int_{\pa\cm}d^dx \sqrt{-\g}\ca\frac1d D_i\xi^i.
\eea
Therefore, if the anomaly vanishes, this charge is conserved for any 
asymptotic conformal Killing vector. However, if the anomaly is non-zero, it 
is only conserved for symmetries associated with boundary Killing vectors.

\subsection{Holographic charges}
\label{holographic_charges}

Let us now derive an alternative form of the conserved charges by
considering instead of (\ref{Noether_variations}) the following variations:
\be\label{Noether_variations_hol}
\d_1'\psi=\cl_{\e\xi}\psi,
\quad\d_2'\psi=\d_\a\psi,
\ee 
where $\xi^i$ is again an asymptotic conformal Killing vector but now $\a$ and
$\e$ reduce to arbitrary functions on $\S_{r_o}$. In contrast to 
(\ref{Noether_variations}), these field variations 
{\em are} a symmetry of the action, but they violate the boundary conditions 
(\ref{boundary_condition}). 

Since these are symmetries of the renormalized action we have
\bea
0=\d_2' I_{\rm ren} & = &\int_{\cm_{r_o}} {\bf E}\d_2'\psi 
+ \int_{\S_{r_o}} d^dx\sqrt{-\g}\p\sub{d}^i\pa_i\a\NO\\
& = &\int_{\cm_{r_o}} {\bf E}\d_2'\psi 
-\int_{\S_{r_o}} d^dx\sqrt{-\g}\a\pa_i\p\sub{d}^i. 
\eea
But now $\a$ is arbitrary and so we conclude that on-shell we must have
\be
\pa_i\p\sub{d}^i=0,
\ee
which also follows immediately from the first equation in (\ref{gf_vector}).
Hence the quantity
\be\label{hol_el_charge}
\cq\equiv-\int_{\pa\cm\cap C}d\s_i\p\sub{d}^i,
\ee
defines a conserved charge, namely the {\em holographic} electric charge.

Similarly,
\bea
0=\d_1' I_{\rm ren} & = & \int_{\cm_{r_o}} {\bf E}\d_1'\psi 
+ \int_{\S_{r_o}} d^dx\sqrt{-\g}\left(\pi\sub{d}^{ij} L_{\e\xi}\g_{ij} 
+ \p\sub{d}^i L_{\e\xi} A_i + \p\sub{\D_I}_I L_{\e\xi}\F^I \right)\NO\\
& = & \int_{\cm_{r_o}} {\bf E}\d_1'\psi 
+ \int_{\S_{r_o}} d^dx\sqrt{-\g}\e\left(\pi\sub{d}^{ij} L_{\xi}\g_{ij} 
+ \p\sub{d}^i L_{\xi} A_i + \p\sub{\D_I}_I L_{\xi}\F^I \right)\NO\\
&&+\int_{\S_{r_o}} d^dx\sqrt{-\g}\left(2\pi\sub{d}^i_j+\p\sub{d}^i A_j\right)
\xi^jD_i\e.
\eea
Therefore, after integration by parts in the last term and using the fact 
that $\e$ is arbitrary, we conclude that on-shell we must have
\bea
D_i\left[(2\pi\sub{d}^i_j+\pi\sub{d}^i A_j)\xi^j\right] & = & \pi\sub{d}^{ij}
L_\xi\g_{ij}+\p\sub{d}^i L_\xi A_i+\pi\sub{\D_I}_I L_\xi\Phi^I\NO\\
& = & \left(2\p\sub{d}^i_i+(\D_I-d)\p\sub{\D_I}_I\F^I\right)\frac1d D_i\xi^i\NO
\\& = & \ca\frac1d D_i\xi^i,
\eea
where we have used the trace Ward identity (\ref{traceWI}) in the last step. 
Hence the quantity 
\beq\label{chargeQ}
\mathcal{Q}[\xi]\equiv\int_{\pa\cm\cap C}d\s_i\left(2\pi\sub{d}^i_j+
\pi\sub{d}^i A_j\right)\xi^j,
\eeq
defines a  holographic conserved charge associated with every asymptotic 
conformal Killing vector, if the anomaly vanishes, or every boundary Killing 
vector, if the anomaly does not vanish. 

From the above analysis we have obtained two apparently different expressions
for the conserved charges associated with every asymptotic symmetry. 
However, we show in the the following lemma that 
the two expressions for the conserved charges are equivalent.

\begin{lemma}\label{hol=noether}
Let $\psi$ denote an AlAdS solution of the bulk equations of motion 
possessing an asymptotic timelike Killing vector $k$ and possibly a 
set of $N-1$ asymptotic spacelike Killing vectors $m_\a$ with closed 
orbits, forming a maximal set of commuting asymptotic isometries.
In adapted coordinates such that $k=\pa_t$ and $m_\a=\pa_{\f^\a}$ the 
background
$\psi$ is independent of the coordinates $x^a=\{t,\f^\a\}$. Then,
\newline
i) \be\label{el_identity}
\int_{\pa\cm\cap C}d\s_i\p\sub{d}^i=\int_{\pa\cm\cap C}\ast\mathcal{F}.
\ee
\newline
ii) If in addition the background metric and gauge field  take asymptotically
the form
\be\label{condition}
d\bar{s}^2\equiv\g_{ij}dx^idx^j=\t_{ab}dx^adx^b+\s_{ij}d\tilde{x}^id
\tilde{x}^j,\quad {\bf A}\equiv A_idx^i=A_adx^a,
\ee
where $\t_{ab}$, $\s_{ij}$ and $A_a$ depend only on the rest of the
transverse coordinates $\tilde{x}^i$ as well as the radial coordinate $r$, 
then, for any asymptotic conformal Killing vector $\x$,
\be
\label{CKV_identity}
\int_{\pa\cm\cap C}d\s_i\left(2\pi\sub{d}^i_j+\pi\sub{d}^i A_j\right)\xi^j=
-\int_{\pa\cm\cap C}\left({\bf Q}[\xi]-i_\xi{\bf B}\right).
\ee
\end{lemma}  

A proof of this lemma can be found in appendix \ref{holographic=Noether}.
However, a few comments are in order regarding the condition 
(\ref{condition}) we have assumed in order to prove the second part of
the lemma. Firstly, as can be seen from the explicit proof, it is only 
required in order to show equivalence of the charges for true
asymptotic {\em conformal} isometries, i.e. with {\em non-zero} conformal 
factor. Otherwise, this condition is not used in the proof. Secondly,
in certain special cases the fact that the background takes the form 
(\ref{condition}) turns out to be a consequence of the existence of the set 
of commuting isometries and the field equations. 

More specifically, the condition that the background takes the form
(\ref{condition}) is closely related to the integrability of the 
$D-N$-dimensional submanifolds orthogonal to $k$ and $m_\a$. In
particular, it was shown in \cite{Wald_book}, Theorem 7.7.1, using 
Frobenius' theorem, that for pure gravity in four dimensions, the 2-planes 
orthogonal to a timelike isometry $k$ and a rotation $m$ are integrable, and 
hence the metric takes the form (\ref{condition}). This result can be
easily extended to include a Maxwell field as well as scalar fields in four 
dimensions \cite{Carter69, Heusler}. More recently, this result was 
generalized for pure gravity in $D$ 
dimensions and $D-2$ orthogonal (non-orthogonal) commuting isometries in
\cite{Emparan&Reall} (\cite{Harmark}).
It appears that these results cannot be generalized in a straightforward way to
include gauge fields for $D>4$, or for less than $D-2$ commuting isometries
in $D$ dimensions. Obviously the restriction to $D-2$ commuting isometries
is too strong for our purposes since even $AdS_D$ only has $[(D+1)/2]$
commuting isometries, which is less than $D-2$ for $D>5$. 

Despite the fact that we lack a general proof of (\ref{condition}) as
a consequence of the presence of the commuting isometries and the 
field equations, this condition is satisfied by a very wide range of
AlAdS spacetimes, including Taub-Nut-AdS and Taub-Bolt-AdS 
\cite{Hawking:1998ct,Awad:2000gg}. It would be very interesting 
to determine  what are the most general conditions so that (\ref{condition})
holds.  
  
\subsection{Wald Hamiltonians}

We now give a third derivation of the conserved charges as `Hamiltonians'
on the covariant phase space 
\cite{Crnkovic&Witten,zuckerman,Lee&Wald,Wald&Zoupas}.
Some results relevant to this section are collected in appendix 
\ref{symplectic}. 

Let $\xi$ be an asymptotic conformal Killing vector and $\a$ an asymptotically
constant gauge transformation, namely
\be
\cl_\xi\psi=\cl_{\hat{\xi}}\psi+\d_{\hat{\a}}\psi+\co(e^{-s_+ r}),
\quad\d_\a\psi=\co(e^{-s^+ r}),
\ee
where $\hat{\a}$, $\hat{\xi}$ and $s^+$ are given in appendix
\ref{KV}.
The `Hamiltonians' which generate these symmetries in phase space
must satisfy Hamilton's equations, which in the covariant phase
space formalism take the form 
\be\label{Hamiltonians}
\d H_\xi=\Om_C(\psi,\d\psi,\cl_\xi\psi),\quad
\d H_\a=\Om_C(\psi,\d\psi,\d_\a\psi),
\ee
where the pre-symplectic form $\Om_C$ is defined in (\ref{symplectic_form}).
The  Hamiltonians exist if these equations can be integrated 
in configuration space to give $H_\xi$ and $H_\a$. As is discussed
in appendix D, the symplectic form is independent of the 
Cauchy surface used to define it if the anomaly vanishes 
or if the variations are associated with boundary Killing 
vectors. It follows that the corresponding Hamiltonians
are conserved, provided the `integration' constant is also independent
of the Cauchy surface. We further discuss this issue below.

Let us first consider $H_\a$ which can be obtained very easily.
Using the result for the symplectic form in (\ref{gauge_exact}) we have
\be
\d H_\a=\int_{\pa\cm\cap C}\d{\bf Q}_\a,
\ee 
and hence, up to a constant,
\be
H_\a=\int_{\pa\cm\cap C}{\bf Q}_\a=-\int_{\pa\cm\cap C}\ast\cf,
\ee
taking $\a\to 1$ asymptotically. Therefore, once again, we have derived
the conserved electric charge.

Consider next $H_\xi$. From (\ref{diffeo_exact}) we have
\be\label{var_diffeo_hamiltonian}
\d H_\xi=\int_{\pa\cm\cap C}\left(\d{\bf Q}[\xi]-i_\xi{\bf \Theta}\right).
\ee
This equation has  a non-trivial integrability condition. Applying a second 
variation and using the commutativity of two variations, 
$\d_1\d_2-\d_2\d_1=0$, we obtain the integrability condition \cite{Wald&Zoupas}
\be
\int_{\pa\cm\cap C}i_\xi\om(\psi,\d\psi_1,\d\psi_2)=0.
\ee 
Since $\xi$ is tangent to $\S_r$, from (\ref{pullback_om_bc}) follows
that this is equivalent to
\be  
\int_{\pa\cm\cap C}d^{d-1}x\xi^t\left\{\d_2(\sqrt{-\g}\mathcal{A})\d_1\s
-1\leftrightarrow 2\right\}=0.
\ee
Therefore, if  the anomaly vanishes, a Hamiltonian associated to any
asymptotic CKV $\xi$ exists.
However, if there is an anomaly and $\xi^t\neq 0$, then a Hamiltonian for 
$\xi$ exists only if the stronger boundary condition (\ref{anbc}) 
is used - i.e. a particular representative of the conformal class is kept 
fixed - in agreement with the analysis of the variational problem.   

The same conclusion can be drawn by trying to integrate 
(\ref{var_diffeo_hamiltonian}) directly. This is possible provided
we can find a $d$-form ${\bf B}$ such that
\be \label{intTh}
\int_{\pa\cm\cap C} i_\xi{\bf \Theta}=\d\int_{\pa\cm\cap C} i_\xi{\bf B}.
\ee
Once such a form is found, then $H_\xi$ exists and it is given by 
\be \label{hami}
H_\xi=\int_{\pa\cm\cap C} ({\bf Q}[\xi]-i_\xi{\bf B}).
\ee
However, since $\xi$ is tangent to $\S_r$, we can use 
(\ref{pullback_theta_exp}), the boundary conditions 
(\ref{boundary_condition}) and the trace Ward identity (\ref{traceWI}) to 
obtain
\be\label{integrate_theta}
\int_{\S_{r_o}\cap C} i_\xi{\bf \Theta}=\frac{1}{\k^2}\d\int_{\S_{r_o}\cap C}
d\s_i\xi^i[K-(K-\l)_{\rm ct}]-\int_{\S_{r_o}\cap C}d\s_i \xi^i\ca\d\s.
\ee 
Therefore, if $\xi^t\neq 0$, then such a form exists for the boundary 
conditions (\ref{boundary_condition}) provided 
the anomaly vanishes, in complete agreement with the conclusion 
from the integrability condition. Moreover, (\ref{integrate_theta}) 
shows that when such a ${\bf B}$ exists it coincides with ${\bf B}$ in 
(\ref{B}) and hence, the corresponding Hamiltonian is precisely the Noether 
charge (\ref{BCKV_Noether_charge}). 

Notice that the Wald Hamiltonians are only defined up to quantities
in the kernel of the variations. In particular, 
when integrating (\ref{intTh}) to obtain (\ref{hami}),
one can add to $H_{\xi}$ an integral of a local density 
constructed only from boundary data and the asymptotic 
conformal Killing vector $\xi$. This `integration constant'
is constrained by the fact that the Hamiltonians
should be conserved. In particular, if $H_\xi$ is a Wald
Hamiltonian, so is 
\be
H_{\xi}' = H_{\xi} + \int_{\pa \cm \cap C} d \s_i H^i_j \xi^j,
\ee
provided $H^i_j$ is constructed locally from boundary data, has
dilatation weight $d$, and it is covariantly conserved. 

In fact such ambiguity is present in $AAdS_{2k+1}$ spacetimes and
has caused some confusion in the literature.  $AAdS_{2k+1}$ spacetimes are 
special in that the boundary is conformally flat, and 
even-dimensional conformally
flat spacetimes admit local covariantly conserved stress 
energy tensors. This is true in all even dimensions, as 
we discuss in appendix \ref{Weyl-tensor}. The best known 
case is the four dimensional one: the tensor
\be \label{h3}
H^i_j\equiv \frac14\left(-R^i_kR^k_j+\frac23RR^i_j+
\frac12R^k_lR^l_k\d^i_j-\frac14R^2\d^i_j\right)
\ee
is covariantly conserved provided the metric is conformally flat.
This tensor is well-known from studies of quantum 
field theories in curved backgrounds, see \cite{BD} (where it 
is called ${}^3H_{\m \n}$).
It has been called `accidentally conserved' in 
\cite{BD} because it is not the limit of a local
tensor that is conserved in non-conformally flat spacetimes
and cannot be derived by varying a local term.
The same tensor is the holographic stress energy tensor 
of\footnote{More precisely, (\ref{h3}) is the 
holographic stress
energy tensor associated to a bulk solution that is conformally flat, see
(3.20) of \cite{dHSS}; 
all such solutions are locally isometric to $AdS_5$.} $AdS_5$ \cite{dHSS}! 

This tensor also appeared recently in comparisons between the 
`conformal mass' of \cite{AD} and the holographic 
mass, see \cite{AD,Hollands:2005wt} and appendix \ref{Weyl-tensor}.
It follows that the conserved charges according to  
both definitions are Wald Hamiltonians. It also follows from this
discussion that the conformal mass is the mass of the spacetime 
{\it relative} to the AdS background. Furthermore, 
we conclude that the definition of \cite{AM} does not extend to general 
AlAdS spacetimes since $H_{ij}$ is not covariantly 
conserved when the  boundary metric is not conformally flat and we already 
know that the holographic charges are conserved for general AlAdS
(and as shown in this section are also Wald Hamiltonians).

\section{The first law of black hole mechanics}
\setcounter{equation}{0}

The above detailed description of the conserved charges allows us to 
study the thermodynamics of AlAdS black hole spacetimes quite generically.  
In particular, we will consider a black hole solution of (\ref{Lagrangian}) 
possessing a timelike Killing vector $k$ and possibly a set of spacelike 
isometries with closed orbit forming a maximal set of commuting isometries as 
in lemma \ref{hol=noether}, but here we require that these isometries be 
exact and not just asymptotic. The form (\ref{condition}) of the metric then 
implies that these bulk isometries correspond to boundary isometries and
not merely asymptotic boundary conformal isometries. Moreover, we 
will assume that the event horizon, $\cn$, of the black hole is a 
non-degenerate bifurcate Killing horizon of a timelike (outside the horizon) 
Killing vector $\chi$ such that the surface gravity, $\hat{\k}$, of the 
horizon is given by
\be
\hat{\k}^2=-\frac12\nabla^\m\chi^\n\nabla_{\m}\chi_\n|_{_\cn}.
\ee   
The inverse temperature, $\b$, then is
\be
\b=T^{-1}=\frac{2\p}{\hat{\k}}.
\ee 

Let us begin with a lemma which is central to our analysis.

\begin{lemma}\label{lemma}\mbox{}
Let $\xi$ be a bulk Killing vector, $I$ the 
renormalized on-shell Euclidean action and $\ch=\cn\cap C$ the
intersection of the horizon with the Cauchy surface. Let also $t$ be the
adapted coordinate to the timelike isometry $k$ so that $k=\pa_t$. 

i) If $\xi^t=1$, then\footnote{Note that the integrals
over $\ch$ should be done with an {\em inward}-pointing unit vector.} 
\be
\b \cq[\xi]-I=-\b\int_{\mathcal{H}}{\bf Q}[\xi].
\ee

ii) If $\xi^t=0$, then 
\be
\cq[\xi]=-\int_{\mathcal{H}}{\bf Q}[\xi].
\ee
\end{lemma}

{\bf Proof:}

By Stokes' theorem\footnote{We assume throughout this paper
that all fields are regular outside and on the horizon
so that the application of Stokes' theorem is legitimate.}
\bea
\int_{\pa\cm\cap C} {\bf Q}[\xi] & = & \int_C {\rm d}{\bf Q}[\xi] 
+\int_{\mathcal{H}}{\bf Q}[\xi]\NO\\
& = &\int_C\left({\bf \Theta}(\psi,\cl_\x\psi)-i_\x{\bf L}\right)
+\int_{\mathcal{H}}{\bf Q}[\xi]\NO\\
& = & -\int_C i_\x{\bf L}+\int_{\mathcal{H}}{\bf Q}[\xi].
\eea
Now, (\ref{lambda}) and the fact that the background is stationary allow us to 
write 
\be
\int_{C} i_\xi{\bf L}=-\int_{\S_{r_o}\cap C}d\s_i\xi^i\l,
\ee
where the minus sign arises due to our choice of orientation (see footnote
\ref{orientation}). Hence,
\bea
\int_{\S_{r_o}\cap C} ({\bf Q}[\xi]-i_\xi{\bf B}) 
& = &  \int_{\mathcal{H}}{\bf Q}[\xi]-\int_{C} i_\xi{\bf L}
-\frac{1}{\k^2}\int_{\S_{r_o}\cap C}d\s_i\xi^i\left(K\sub{d}
+\l_{\rm ct}\right)\NO\\
& = & \int_{\mathcal{H}}{\bf Q}[\xi]-\frac{1}{\k^2}\int_{\S_{r_o}}d\s_i
\xi^i\left(K\sub{d}-\l\sub{d}\right).
\eea
For $\xi^t=0$ the last term vanishes. If however
$\xi^t=1$, then we can use the fact that the background is
stationary to obtain
\be
\frac{\b}{\k^2}\int_{\pa\cm\cap C}d\s_i\xi^i\left(K\sub{d}-\l\sub{d}
\right)=-\frac{1}{\k^2}\int_{\pa\cm}d^dx\sqrt{\g_E}\left(K\sub{d}-\l\sub{d}
\right)\equiv I,
\ee
where $\g_{Eij}$ is the Euclidean metric.
This completes the proof.
\begin{flushright}
$\square$
\end{flushright}

\subsection{Black hole thermodynamics}

This lemma, besides relating the conserved charges to local integrals over 
the horizon, leads immediately to the quantum statistical relation
\cite{GH}
\be\label{qm_relation} 
I=\b G\left(T,\Om_i,\F\right),
\ee
where
\be
G\left(T,\Om_i,\F_i\right)\equiv M-TS-\Om_iJ_i-\F\, Q,
\ee 
is the Gibbs free energy. (\ref{qm_relation}) follows trivially from 
lemma \ref{lemma} provided 
\be\label{Gibbs_id}
\cq[\chi]+\int_{\mathcal{H}}{\bf Q}[\chi]=  M-TS-\Om_iJ_i-\F\,Q,
\ee
where $\chi$ is the null generator of the horizon, normalized such
that $\chi^t=1$. To show that this is the case we 
need a precise definition of the thermodynamic variables appearing in 
the Gibbs free energy.

\begin{flushleft}
{\em Electric charge}
\end{flushleft} 

Using Stokes' theorem, the electric charge\footnote{
We assume in this paper that the black holes are only electrically charged.
If there are magnetic charges as well, one has to be careful 
with global issues.} (\ref{elch}) is also given by
\be
Q\equiv -\int_{\pa\cm\cap C}\ast\cf=-\int_{\ch}\ast\cf.
\ee

\begin{flushleft}
{\em Electric potential}
\end{flushleft} 

We define the electric potential, $\F$, conjugate to the charge $Q$, by
\be
\F\equiv-A_\m\chi^\m|_\ch.
\ee
This is well-defined, for $A_\m\chi^\m$ is constant on $\ch$. To see this 
consider a vector field $t$ tangent to the horizon. Then,
\be
t\cdot\pa(A_\m \chi^\m)=t^\r(\chi^\m F_{\r\m}+\cl_\chi A_\r)=t^\r\chi^\m 
F_{\r\m}.
\ee  
But since $t$ is tangent to $\ch$, $t|_{_\ch}\propto \chi$ and hence
\be
t\cdot\pa(A_\m \chi^\m)|_{_\ch}=0.
\ee

\begin{flushleft}
{\em Mass}
\end{flushleft} 

In order to define the mass we have to supply an asymptotic timelike
Killing vector. In contrast to asymptotically flat spacetimes, 
in AlAdS spacetimes there is an additional subtlety in that there
can be a non-zero angular velocity, $\Om^\infty_i$,  at spatial infinity.
This is the case, for example, for the Kerr-AdS black holes in 
Boyer-Lindquist coordinates as we will see below. In such a rotating
frame, there are many timelike Killing vectors obtained by 
appropriate linear combinations of $\pa_t$ and $\pa_{\phi_i}$.
Using a general timelike Killing vector will result
in a conserved quantity that is a linear combination of the 
true mass and the angular momenta. To resolve this issue 
we first go to a non-rotating frame by the coordinate transformation
\be
t'=t,\quad \f_i'=\f_i-\Om^\infty_i t.
\ee
In this frame there is no such ambiguity and one can 
define the mass, as usual, using the Killing vector $\pa_{t'}$.
In terms of the original coordinates we have
\be\label{timelike_kv}
\pa_{t'}=\frac{\pa t}{\pa t'}\pa_{t}+\frac{\pa\f_i}{\pa t'}\pa_{\f_i}=
\pa_{t}+\Om^\infty_i\pa_{\f_i}.
\ee
Therefore, the mass is defined as
\be\label{mass}
M\equiv \cq[\pa_{t}+\Om^\infty_i\pa_{\f_i}].
\ee

\begin{flushleft}
{\em Angular velocities}
\end{flushleft} 

Let $\chi = \pa_t+\Om_i^H\pa_{\f_i}$ be the null generator of the 
horizon. This defines the angular velocities, $\Om_i^H$, of the horizon. 
We define the angular velocities, $\Om_i$,  by
\be\label{angular_velocity}
\Om_i\equiv \Om_i^H-\Om^\infty_i.
\ee

\begin{flushleft}
{\em Angular momenta}
\end{flushleft} 

We define the angular momenta, $J_i$, by
\be
J_i\equiv -\cq[\pa_{\f_i}]=\int_{\mathcal{H}}{\bf Q}[\pa_{\f_i}],
\ee 
where the second equality follows from lemma \ref{lemma}.

\begin{flushleft}
{\em Entropy}
\end{flushleft} 

Finally, using Wald's definition of the entropy \cite{Wald93} 
(see also \cite{Jacobson:1993vj}) we get
\be\label{entropy}
-\b\int_{\mathcal{H}}{\bf Q}[\chi]=S+\b \F\,Q. 
\ee
 
With these definitions it is now straightforward to see that (\ref{Gibbs_id}),
and hence (\ref{qm_relation}) hold.

\subsection{First law}

To derive the first law we consider variations that satisfy
our boundary conditions.  Namely, if the anomaly
vanishes, then the boundary conditions (\ref{boundary_condition}) should
be satisfied, otherwise (\ref{anbc}) should hold, i.e. a representative 
of the conformal class 
should be kept fixed. We will discuss the significance of this in the 
next section. In other words, we only vary the normalizable 
mode of the solutions, as one might have anticipated on physical 
grounds.\footnote{\label{var_norm} Note that the non-normalizable mode 
determines the conformal class at the boundary. The non-normalizable mode 
{\em together} with a defining function specify a representative of the 
conformal class.}

We now show, following Wald et al. \cite{Wald93,Iyer:1994ys}, that 
these variations satisfy the first law.
From equation (\ref{diffeo_exact}) we have
\be
{\rm d}\left(\d{\bf Q}[\chi]-i_\chi {\bf \Theta}\right)=
\om(\psi,\d\psi,\cl_\chi\psi).
\ee
Hence,
\bea
\int_C{\rm d}\left(\d{\bf Q}[\chi]-i_\chi {\bf \Theta}\right) & = &
\int_{\pa\cm\cap C}\left(\d{\bf Q}[\chi]-i_\chi {\bf \Theta}\right)
-\int_{\ch}\left(\d{\bf Q}[\chi]-i_\chi {\bf \Theta}\right)\NO\\
& = & \int_C\om(\psi,\d\psi,\cl_\chi\psi)=0,
\eea
since $\chi$ is a Killing vector. However, $\chi$ is tangent 
to $\ch$ and so we arrive at
\be\label{first_law_id}
\int_{\pa\cm\cap C}\left(\d{\bf Q}[\chi]-i_\chi {\bf \Theta}\right)
=\int_{\ch}\d{\bf Q}[\chi].
\ee 
Consider first the left hand side. Writing $\chi=\pa_t+\Om^\infty_i\pa_{\f_i}
+\Om_i\pa_{\f_i}$ and using the fact that $\pa_{\f_i}$ is tangent to 
$\pa\cm$ we get
\bea\label{first_law_der}
\int_{\pa\cm\cap C}\left(\d{\bf Q}[\chi]-i_\chi {\bf \Theta}\right) & = &
\int_{\pa\cm\cap C}\left(\d{\bf Q}[\pa_t+\Om^\infty_i\pa_{\f_i}]
-i_t{\bf \Theta}\right)+\Om_i\int_{\pa\cm\cap C}\d{\bf Q}[\pa_{\f_i}]\NO\\
& = &
\d\int_{\pa\cm\cap C}\left({\bf Q}[\pa_t+\Om^\infty_i\pa_{\f_i}]
-i_t{\bf B}\right)+\Om_i\d\int_{\pa\cm\cap C}{\bf Q}[\pa_{\f_i}]\NO\\
& = & - (\d M-\Om_i\d J_i).
\eea
In order to evaluate the right hand side  
of (\ref{first_law_id}) we need to match the
horizons of the perturbed and unperturbed solutions \cite{Wald93}, 
the unit surface gravity generators, $\tilde{\chi}\equiv 
\frac{1}{\hat{\k}}\chi$, of the horizons and the electric potentials.
From (\ref{entropy}) then we immediately get
\be
-\int_\ch \d{\bf Q}[\chi]=T\d S+\F \d Q.
\ee

Therefore, (\ref{first_law_id}) is a statement of the first law, namely
\be\label{first_law}
\d M=T\d S+\Om_i\d J_i+\F \d Q.
\ee

However, we emphasize that  the variations in this expression must satisfy 
the appropriate boundary conditions that make the variational problem 
well-defined. Namely, if the anomaly
vanishes, then the boundary conditions (\ref{boundary_condition}) should
be satisfied, but if there is a non-zero anomaly, then  (\ref{anbc}) 
must be satisfied instead, i.e. the representative of the conformal class 
should be kept fixed. We will discuss the significance of this in the 
next section.

\subsection{Dependence on  the representative of the conformal class}

Let us now discuss how the thermodynamic variables
defined above depend on the representative of the conformal class at the 
boundary. 

To this end we recall that a Weyl transformation on the boundary is induced
by a PBH transformation, i.e. a combined  bulk diffeomorphism and a 
compensating gauge transformation, given by (\ref{PBH_transformation})
after setting $\xi_o=0$ and $\a_o=0$. However, in order to be able to 
compare the mass and angular momenta for the two representatives of the 
conformal class we require that  the two representatives have the
same maximal set of commuting isometries, i.e. we restrict to Weyl 
factors $\d\s$ which are independent of the coordinates $t,\;\f_\a$ adapted 
to the asymptotic isometries.

It is now straightforward to see that all intensive thermodynamic variables,
namely the temperature $T$, the angular velocities $\Om_i$ and the
electric potential $\F$ are invariant under such diffeomorphisms.
The same holds for the entropy $S$, the angular momenta $J_i$, and
the electric charge $Q$, since, as we saw above, they can be expressed as
local integrals over the horizon. Therefore, the only quantities which could
potentially transform non-trivially are the mass $M$ and the on-shell
Euclidean action $I$. However, their transformations are not independent since 
they are constrained by the quantum statistical relation (\ref{qm_relation}),
namely
\be\label{Weyl}
\d_\s I=\b\d_\s M.
\ee  
This is a significant result which cannot be seen easily otherwise. 
We know that 
\be
\d_\s I=-\int_{\pa\cm}d^dx\sqrt{\g_E}\ca\d\s,
\ee
while
\be
\d_\s M=-2\int_{\pa\cm\cap C}d\s_i\{(2\tilde{\p}\sub{d}^i_j+
\tilde{\p}\sub{d}^iA_j)\tilde{k}^j\d\s+\ldots\},
\ee
where $\tilde{k}=\pa_t+\Om_i^\infty\pa_{\f_i}$ and the dots stand for terms
involving derivatives of the Weyl factor $\d\s$.  One can check this 
explicitly in certain examples by directly evaluating the transformation
of the renormalized stress tensor under a PBH transformation 
\cite{dHSS,Skenderis_proceedings}.

As a final point let us consider how (\ref{first_law}) would be
modified if there is a non-vanishing anomaly and we allow for
variations which keep fixed only the conformal class and not a 
particular representative. In this case, (\ref{integrate_theta})
implies that 
\be
-\int_{\pa\cm\cap C}\left(\d{\bf Q}[\chi]-i_\chi {\bf \Theta}\right)
= - T\d_\s I + \d M - \Omega_i \delta J_i,
\ee
and the first law should be modified to 
\bea\label{first_law_modified}
\d M & = & T\d_\s I+T\d S+\Om_i \d J_i +\F \d Q\NO\\
& = &  \d_\s M+T\d S+\Om_i \d J_i +\F \d Q,
\eea
where $\d\s$ is the Weyl factor by which the representative of the 
conformal class is changed due to the variation $\d$ and the second
equality follows from (\ref{Weyl}). 

We can now state precisely how the first law works in the presence
of a non-vanishing anomaly. A generic variation $\d$ will not keep
the conformal representative fixed and it will induce a Weyl transformation
$\d\s$. We should then undo this Weyl transformation by a PBH 
transformation with Weyl factor $-\d\s$. Then, (\ref{first_law_modified})
ensures that the combined variation, which {\em does} keep the conformal 
representative fixed, satisfies the usual first law. The general
Kerr-AdS black hole in five dimensions provides a clear illustration
of this.

\section{Examples}
\setcounter{equation}{0}

In this section we will demonstrate our analysis by two examples,
the Kerr-Newman-AdS black hole in four dimensions \cite{Carter68,
Plebanski} and the general Kerr-AdS black hole in five dimensions 
\cite{HHTR}. The second example provides a clear illustration of the
role of the conformal anomaly in the thermodynamics. 

Before focusing on the specific examples however we discuss 
the steps and subtleties involved in the computation. 
Recall that the defining feature of the counterterm method is that
the on-shell action of AdS gravity can be rendered finite
on {\em any} solution by adding to the action a 
set of {\em local covariant} boundary counterterms.
One should not forget, however, that the precise form 
of the counterterms depends on the regularization/renormalization
scheme. The counterterms used in the literature
were derived  using as regulator a cut-off in the 
Fefferman-Graham radial coordinate $z$ \cite{HS}, or equivalently in the
radial coordinate $r$ we use in this paper. The cut-off hypersurface
$r=r_o$ is in general different from the hypersurfaces $\tilde{r}=const.$,
where $\tilde{r}$ is another radial coordinate that might appear 
naturally in the bulk metric. 
So, to evaluate correctly the counterterm contribution to the 
on-shell action, one should transform asymptotically
the solution to Fefferman-Graham coordinates and then evaluate the 
counterterm action (or equivalently transform the hypersurface $r=r_o$ and 
the counterterm action in the new coordinates). Of course, it is always 
possible to work with a different regulator but then the counterterm
action should be worked out from scratch.

Let us discuss now the evaluation of the conserved charges.
Given that the asymptotics and counterterms are universal,
one can work out in full generality the explicit form of the 
renormalized stress energy tensor in terms of the metric coefficients 
$\mathtt{g}_{(m)}$ that appear in the asymptotic expansion
of the solutions of a given action. This is done 
for pure gravity in \cite{dHSS} 
and for gravity coupled to certain matter in \cite{holren}. To evaluate 
the holographic stress tensor on a specific solution 
one could thus simply read off the metric 
coefficients from the asymptotic expansion of the metric
and plug them in the general formula. This is the simplest 
way to proceed if the explicit expression for the holographic 
stress energy tensor is known. If this is not the case, it is simpler 
to just compute from the asymptotics of the given solution 
the contribution of the bulk 
and counterterm actions to the holographic stress energy tensor 
and add them up to produce a finite answer. To evaluate the 
conserved charges we finally integrate the holographic 
stress energy tensor contracted with the appropriate asymptotic
conformal Killing vector over the appropriate domain. 
The only remaining subtlety 
is the choice of a timelike Killing vector to be used in the definition
of mass when the boundary metric is in a rotating frame. 
In this case we choose the Killing vector that corresponds to the 
standard timelike Killing vector $\pa/\pa t$ is the 
corresponding non-rotating frame.

Below we describe our calculation for the four-dimensional Kerr-Newman-AdS 
black hole in considerable detail, mainly in order to emphasize the role of the
Fefferman-Graham coordinate system in the asymptotic analysis, which 
is not fully appreciated in the literature.  We then turn to the 
five dimensional Kerr-AdS black hole, emphasizing the role of the 
anomaly and its relation to the Casimir energy. Previous work 
on the thermodynamics of these black holes includes
\cite{Henneaux:1984xu, Kostelecky:1995ei,HHTR,Das&Mann, 
Silva:2002jq, Caldarelli,Awad&Johnson,GPP,Deruelle,Barnich&Compere,
Olea:2005gb}. 

\subsection{D=4 Kerr-Newman-AdS black hole}

The metric of the Kerr-Newman-AdS black hole in Boyer-Lindquist 
coordinates reads \cite{Carter68,Plebanski,Caldarelli}
\be\label{metric_1}
ds^2=-\frac{\D_r}{\r^2}\left(dt-\frac{a\sin^2\theta}{\Xi}d\f\right)^2+
\frac{\r^2}{\D_r}dr^2+\frac{\r^2}{\D_\th}d\th^2+
\frac{\D_\th\sin^2\th}{\r^2}\left(adt-\frac{r^2+a^2}{\Xi}d\f\right)^2,
\ee
where
\bea
&&\r^2=r^2+a^2\cos^2\th,\NO\\
&&\D_r=(r^2+a^2)\left(1+\frac{r^2}{l^2}\right)-2mr+q^2,\NO\\
&&\D_\th=1-\frac{a^2}{l^2}\cos^2\th,\quad\X=1-\frac{a^2}{l^2}.
\eea
The gauge potential in this coordinate system is given by
\be
{\bf A}=-\frac{2qr}{\r^2}\left(dt-\frac{a\sin^2\theta}{\Xi}d\f\right).
\ee
This metric and gauge field solve the Einstein-Maxwell equations 
which follow from the action (omitting the boundary terms) 
\be
I_{\rm Lorentzian}=\frac{1}{2\k^2}\int_\cm d^4x\sqrt{-g}\left(R-2\L-
\frac14 F^2\right).
\ee

The event horizon is located at $r=r_+$, where $r_+$ is the largest root
of $\D_r=0$, and its area is
\be
A=\frac{4\p(r_+^2+a^2)}{\X}.
\ee
The analytic continuation of  the Lorentzian metric (\ref{metric_1}) by
$t=-i\t$, $a=i\a$ develops a conical singularity unless we periodically
identify $\t\sim\t+\b$ and $\f\sim\f+i\b\Om_H$, where 
\be
\b=\frac{4\p(r_+^2+a^2)}{r_+\left(1+\frac{a^2}{l^2}+\frac{3r_+^2}{l^2}-
\frac{a^2+q^2}{r_+^2}\right)},
\ee
is the inverse temperature and the angular velocity of the horizon, 
$\Om_H$, is given by
\be
\Om_H=\frac{a\X}{r_+^2+a^2}.
\ee
However, in this coordinate system there is a non-zero angular velocity at
infinity, namely
\be
\Om_\infty=-\frac{a}{l^2}.
\ee
Following our prescription (\ref{angular_velocity}), we define the angular 
velocity relevant for the thermodynamics as the difference
(see also \cite{Caldarelli,GPP})
\be
\Om=\Om_H-\Om_\infty=\frac{a(1+r_+/l^2)}{r_+^2+a^2}.
\ee
Finally, if $\chi=\pa_t+\Om_H\pa_\phi$ is the null generator of the Killing 
horizon, the electric potential is given by
\be
\F\equiv -A_\m\chi^\m|_{r_+}=\frac{2qr_+}{r_+^2+a^2}.
\ee

Next we determine the electric charge, angular momentum and mass, as well as 
the Euclidean on-shell action of the Kerr-Newman-AdS solution.
Our general analysis of the charges in section \ref{conserved_charges} 
showed that the counterterms do not contribute to the value of the electric 
charge or the angular momentum (lemma \ref{hol=noether}). 
However, the counterterms are essential for evaluating the
mass and the on-shell action. Starting with the electric charge we
easily find
\be
Q\equiv-\frac{1}{2\k^2}\int_{\pa\cm\cap C}\ast {\rm d}{\bf A}=
\frac{4\p q}{\k^2\X}.
\ee
The angular momentum can be evaluated equally easily as
\be\label{J}
J\equiv \int_{\pa\cm\cap C}{\bf Q}[\pa_\f]=\frac{8\p ma}{\k^2 \X^2}.
\ee

Before we can calculate  the mass and the on-shell Euclidean action, we 
must first carry out the asymptotic analysis and determine the counterterms.
Expanding the metric (\ref{metric_1}) for large $r$ we get
\bea\label{metric_1_exp}
ds^2 & = & -\frac{r^2}{l^2}\left[1+\left(1+\frac{a^2}{l^2}\sin^2\th\right)
\frac{l^2}{r^2}-\frac{2ml^2}{r^3}+\co\left(\frac{1}{r^4}\right)\right]
\left(dt-\frac{a\sin^2\theta}{\Xi}d\f\right)^2\NO\\
&&+\frac{l^2}{r^2}\left[1-\left(1+\frac{a^2}{l^2}\sin^2\th\right)
\frac{l^2}{r^2}+\frac{2ml^2}{r^3}+\co\left(\frac{1}{r^4}\right)\right]dr^2\NO\\
&&+\frac{r^2}{\D_\th}\left(1+\frac{a^2}{r^2}\cos^2\th\right)d\th^2\NO\\
&&+\frac{r^2\D_\th\sin^2\th}{\X^2}\left[d\f^2+\frac{a^2}{r^2}\left(
(1+\sin^2\th)d\f^2-\frac{2\X}{a}d\f dt\right)+\co\left(\frac{1}{r^4}\right)
\right].
\eea 
This metric is not of the standard form 
since the coefficient of the radial line element
depends on the angle $\theta$. Indeed the standard counterterms are
derived using a Fefferman-Graham coordinate system of the form
(\ref{gf_metric}) \cite{HS,dHSS, Skenderis:2002wp,PS1}. These 
counterterms, defined on hypersurfaces of {\em constant Fefferman-Graham
radial coordinate}, are not necessarily the correct counterterms 
on the hypersurfaces of constant Boyer-Lindquist radial coordinate,
as is widely assumed in the literature. Of course, it is in principle
possible to choose a gauge which asymptotes to the Boyer-Lindquist
form of the Kerr-AdS metric and carry out the asymptotic analysis from 
scratch using a regulator of constant Boyer-Lindquist radial coordinate
and rederive the appropriate counterterms for such a regulator. 
However, it is much more efficient to bring the metric (\ref{metric_1_exp})
into the Fefferman-Graham form and use the standard counterterms.

To this end we introduce new coordinates
\bea
&&\bar{r}=r+\frac1r f(\th)+\co\left(\frac{1}{r^3}\right),\NO\\
&&\bar{\th}=\th+\frac{1}{r^4} h(\th)+\co\left(\frac{1}{r^6}\right),
\eea
or
\bea
&& r=\bar{r}\left[1-\frac{1}{\bar{r}^2} f(\bar{\th})+\co\left(\frac{1}
{\bar{r}^4}\right)\right],\NO\\
&&\th=\bar{\th}-\frac{1}{\bar{r}^4} h(\bar{\th})+\co\left(\frac{1}{\bar{r}^6}
\right).
\eea
Requiring that the coefficient of the new radial line element has no
angular dependence and that there is no mixed term $d\bar{r}d\bar{\th}$ in 
the metric  uniquely fixes the functions $f(\bar{\th})$ and $h(\bar{\th})$ 
to be
\bea
&& f(\bar{\th})=-\frac{a^2}{4}\cos^2\bar{\th},\NO\\
&& h(\bar{\th})=\frac18l^2a^2\D_{\bar{\th}}\sin\bar{\th}\cos\bar{\th}.
\eea
In the new coordinate system the asymptotic form of the metric (\ref{metric_1})
becomes
\bea
ds^2 & = & \frac{l^2}{\bar{r}^2}\left[1-\left(1+\frac{a^2}{l^2}\right)
\frac{l^2}{\bar{r}^2}+\frac{2ml^2}{\bar{r}^3}+\co\left(\frac{1}{\bar{r}^4}
\right)\right]d\bar{r}^2\NO\\
&&+\frac{\bar{r}^2}{\D_{\bar{\th}}}\left[1+\frac32\frac{a^2}{\bar{r}^2}
\cos^2\bar{\th}+\co\left(\frac{1}{\bar{r}^4}\right)\right]d\bar{\th}^2\NO\\
&&-\frac{\bar{r}^2}{l^2}\left[1+\left(1+\frac{a^2}{l^2}-\frac{a^2}{2l^2}
\cos^2\bar{\th}\right)\frac{l^2}{\bar{r}^2}-\frac{2ml^2}{\bar{r}^3}+
\co\left(\frac{1}{\bar{r}^4}\right)\right]
\left(dt-\frac{a\sin^2\bar{\th}}{\Xi}d\f\right)^2\NO\\
&&+\frac{\bar{r}^2\D_{\bar{\th}}\sin^2\bar{\th}}{\X^2}\left[d\f^2+
\frac{a^2}{\bar{r}^2}\left(
(2-\frac12\cos^2\bar{\th})d\f^2-\frac{2\X}{a}d\f dt\right)
+\co\left(\frac{1}{\bar{r}^4}\right)\right],
\eea 
which is almost of the desired form. 
For later convenience let us write explicitly
the components of the induced metric:
\bea\label{metric_1_FG}
&&\g_{\bar{\th}\bar{\th}}=\frac{\bar{r}^2}{\D_{\bar{\th}}}\left[1+
\frac32\frac{a^2}{\bar{r}^2}\cos^2\bar{\th}+\co\left(\frac{1}{\bar{r}^4}
\right)\right],\NO\\
&&\g_{tt}=-\frac{\bar{r}^2}{l^2}\left[1+\left(1+\frac{a^2}{l^2}-\frac{a^2}
{2l^2}\cos^2\bar{\th}\right)\frac{l^2}{\bar{r}^2}-\frac{2ml^2}{\bar{r}^3}+
\co\left(\frac{1}{\bar{r}^4}\right)\right],\NO\\
&&\g_{t\f}=\frac{\bar{r}^2a\sin^2\bar{\th}}{l^2\X}\left[1+\left(1+\frac12
\cos^2\bar{\th}\right)\frac{a^2}{\bar{r}^2}-\frac{2ml^2}{\bar{r}^3}+
\co\left(\frac{1}{\bar{r}^4}\right)\right],\NO\\
&&\g_{\f\f}=\frac{\bar{r}^2\sin^2\bar{\th}}{\X}\left[1+\left(1+\frac12
\cos^2\bar{\th}\right)\frac{a^2}{\bar{r}^2}+\frac{2ma^2\sin^2\bar{\th}}
{\bar{r}^3\X}+\co\left(\frac{1}{\bar{r}^4}\right)\right].
\eea

We can now introduce a cut-off at 
$\bar{r}=\bar{r}_o$ and proceed with the asymptotic analysis in the
standard fashion. Note that the regulating surface $\bar{r}=\bar{r}_o$
becomes angle-dependent in the Boyer-Lindquist coordinates, namely
\be
r_o(\th)=\bar{r}_o\left[1+\frac{a^2}{4\bar{r}_o^2}\cos^2\th+
\co\left(\frac{1}{\bar{r}_o^4}\right)\right].
\ee
This is precisely the reason why the counterterms on a regulating 
surface defined by $r_o={\rm constant}$ are not necessarily the
same as the counterterms on a regulating surface defined by 
$\bar{r}_o={\rm constant}$.

Finally, to bring the metric in the form (\ref{gf_metric})
we define the canonical radial coordinate
\be
dr_*=l\left[1-\frac12\left(1+\frac{a^2}{l^2}\right)\frac{l^2}{\bar{r}^2}+
\frac{ml^2}{\bar{r}^3}+\co\left(\frac{1}{\bar{r}^4}\right)\right]
\frac{d\bar{r}}{\bar{r}}.
\ee

\begin{flushleft}
{\em Counterterms}\footnote{We give the counterterms for the Euclidean action 
which we want to evaluate. The counterterms for the Lorentzian action are
easily obtained by analytic continuation.}
\end{flushleft}

Following the standard algorithm for the asymptotic analysis we find
that the counterterm action for the Maxwell-AdS gravity system in
four dimensions is 
\be
I_{\rm ct}=\frac{1}{\k^2}\int_{\S_{\bar{r}_o}}d^3x\sqrt{\g_E}\left(
\frac2l+\frac{l}{2}R\right).
\ee

\begin{flushleft}
{\em On-shell Euclidean action}
\end{flushleft}

We are now ready to evaluate the renormalized on-shell Euclidean action
\be
I_{\rm ren}=-\frac{1}{2\k^2}\int_{\cm_{\bar{r}_o}}d^4x\sqrt{g_E}
\left(R[g_E]+\frac{6}{l^2}-\frac14 F^2\right)
-\frac{1}{\k^2}\int_{\S_{\bar{r}_o}}d^3x\sqrt{\g_E}\left(K
-\frac2l-\frac{l}{2}R\right).
\ee
Since the background is stationary, the bulk integral gives
\bea
&&\frac{\b}{2\k^2}\int_0^{2\p}
d\f\int_0^\p d\th\int_{r_+}^{r_o(\th)}dr
\sqrt{g_E}\left(\frac{6}{l^2}+\frac14 F^2\right)=\NO\\
&&\quad\quad\frac{4\p\b}{\k^2l^2\X}\left[\bar{r}_o\left(\bar{r}_o^2+
\frac54a^2\right)-r_+(r_+^2+a^2)-\frac{q^2l^2r_+}{r_+^2+a^2}+
\co\left(\frac{1}{\bar{r}_o}\right)\right].
\eea
Moreover, the boundary term is
\bea
-\frac{1}{\k^2}\int_{\S_{\bar{r}_o}}d^3x\sqrt{\g_E}\left(K-\frac2l-\frac{l}{2}
R\right)=
-\frac{4\p\b}{\k^2l^2\X}\left[\bar{r}_o(\bar{r}_o^2+\frac54a^2)
-ml^2+\co\left(\frac{1}{\bar{r}_o}\right)\right].
\eea
Hence,
\be
I_{\rm ren}=\frac{4\p\b}{\k^2l^2\X}\left[ml^2-r_+(r_+^2+a^2)-\frac{q^2l^2r_+}
{r_+^2+a^2}\right].
\ee

\begin{flushleft}
{\em Renormalized stress tensor and conserved charges}
\end{flushleft}

We need now to evaluate the renormalized stress tensor
\be
T\sub{3}^i_j  =  -\frac{l}{\k^2}\left(K\sub{3}^i_j-K\sub{3}\d^i_j\right).
\ee
This can be done either by first writing the renormalized stress tensor
in terms of the coefficients in the Fefferman-Graham expansion of
the metric \cite{dHSS,holren,PS1} and then reading off the coefficients from
(\ref{metric_1_FG}), or by first evaluating the extrinsic curvature using
\be
K_{ij}=\frac12\frac{d\bar{r}}{dr_*}\frac{d}{d\bar{r}}\g_{ij},
\ee
and then subtracting the appropriate counterterms, namely 
\be
T\sub{3}^i_j = -\frac{l}{\k^2}\left(K^i_j-K\d^i_j+\frac2l\d^i_j-lR^i_j+
\frac12lR\d^i_j\right)+\co\left(\frac{1}{\bar{r}_o^4}\right).
\ee
In any case we find (in agreement with \cite{Caldarelli})
\bea
&&T\sub{3}^t_t=-\frac{2m}{\k^2}\frac{l^3}{\bar{r}_o^3}+\co\left(\frac{1}
{\bar{r}_o^4}\right),\NO\\
&&T\sub{3}^{\bar{\th}}_{\bar{\th}}=T\sub{3}^{\f}_{\f}=\frac{m}{\k^2}\frac{l^3}
{\bar{r}_o^3}+\co\left(\frac{1}{\bar{r}_o^4}\right),\NO\\
&&T\sub{3}^t_{\f}=\frac{3m}{\k^2}\frac{a\sin^2\bar{\th}}{l\X}\frac{l^3}
{\bar{r}_o^3}+\co\left(\frac{1}{\bar{r}_o^4}\right),\NO\\
&&T\sub{3}^{\f}_t=\co\left(\frac{1}{\bar{r}_o^4}\right).
\eea

For this solution one can easily show that the gauge field momentum does
not contribute to the holographic charge (\ref{chargeQ}) and so, for any 
boundary conformal Killing vector, $\xi$, we have 
\be
\mathcal{Q}[\xi]  =  -\int_0^{2\p}d\f\int_0^\p d\bar{\th}\sqrt{-\g}
T\sub{3}^t_j\xi^j.
\ee
As a check, we evaluate 
\be
\mathcal{Q}[-\pa_\f]=\frac{8\p ma}{\k^2 \X^2},
\ee
in complete agreement with (\ref{J}). 

To obtain the mass now we first need to identify the correct timelike Killing 
vector. This can be done unambiguously as follows. From the asymptotic
form of the metric in Boyer-Lindquist coordinates we see that the 
corresponding boundary metric is not the standard metric on
$\mathbf{R}\times\mathbf{S}^2$ even for $m=q=0$, since there is a non-zero
angular velocity $\Om_\infty=-\frac{a}{l^2}$. However, as it is discussed
e.g. in \cite{GPP}, this boundary metric is {\em conformal} to the
standard boundary metric of $AdS_4$. To see this we perform
a coordinate transformation from the coordinates $(t,\bar{\th},\f)$ to
$(t',\bar{\th}',\f')$, given by
\be
t'=t,\quad\f'=\f+\frac{a}{l^2}t,\quad\X\tan^2\bar{\th}'=\tan^2\bar{\th}.
\ee  
The resulting boundary metric in the new coordinates is the standard
metric on $\mathbf{R}\times\mathbf{S}^2$ up to the conformal factor
$\cos^2\bar{\th}/\cos^2\bar{\th}'$. It follows that the correct
timelike Killing vector that defines the mass is
\be
\pa_{t'}=\frac{\pa t}{\pa t'}\pa_t+\frac{\pa\f}{\pa t'}\pa_\f
=\pa_t-\frac{a}{l^2}\pa_\f,
\ee
in agreement with (\ref{timelike_kv}). Hence,
\be
M\equiv \cq[\pa_t-\frac{a}{l^2}\pa_\f]=\frac{8\p m}{\k^2\X^2}.
\ee
This is precisely the mass obtained in \cite{GPP} by integrating the 
first law.\footnote{Note that our timelike Killing vector is 
{\em different} from the Killing vector, $\pa_t+\frac{a}{l^2}\pa_\f$, 
which the authors of \cite{GPP} 
claim makes the conformal mass \cite{AM,AD,Das&Mann}
equal to the mass obtained from the first law.} 
Finally, defining the entropy by
\be
S=\frac{2\p}{\k^2}A,
\ee
it can now be easily seen that the quantum statistical relation 
(\ref{qm_relation}) as well as the first law (\ref{first_law}) are 
satisfied.

\subsection{D=5 Kerr-AdS black hole}

As a second example we consider the general five dimensional Kerr-AdS 
solution \cite{HHTR}, which illustrates the role of the conformal anomaly. 
 
In Boyer-Lindquist coordinates the metric is\footnote{Note that 
$0\leq\th\leq\pi/2$ in five dimensions, while $0\leq\th\leq\pi$ in four 
dimensions.} 
\bea\label{metric_2}
ds^2 & = & -\frac{\D_r}{\r^2}\left(dt-\frac{a\sin^2\theta}{\Xi_a}d\f
-\frac{b\cos^2\theta}{\Xi_b}d\psi\right)^2+
\frac{\D_\th \sin^2\th}{\r^2}\left(adt-\frac{r^2+a^2}{\Xi_a}d\f\right)^2\NO\\
&&+\frac{\D_\th \cos^2\th}{\r^2}\left(bdt-\frac{r^2+b^2}{\Xi_b}d\psi\right)^2+
\frac{\r^2}{\D_r}dr^2+\frac{\r^2}{\D_\th}d\th^2\NO\\
&&+\frac{(1+r^2l^{-2})}{r^2\r^2}\left(abdt-\frac{b(r^2+a^2)\sin^2\th}{\Xi_a}
d\f-\frac{a(r^2+b^2)\cos^2\th}{\Xi_b}d\psi\right)^2,
\eea
where
\bea
&&\r^2=r^2+a^2\cos^2\th+b^2\sin^2\th,\NO\\
&&\D_r=\frac{1}{r^2}(r^2+a^2)(r^2+b^2)\left(1+\frac{r^2}{l^2}\right)-2m,\NO\\
&&\D_\th=1-\frac{a^2}{l^2}\cos^2\th-\frac{b^2}{l^2}\sin^2\th,\NO\\
&&\X_a=1-\frac{a^2}{l^2},\quad \X_b=1-\frac{b^2}{l^2}.
\eea

The event horizon is located at $r=r_+$, where $r_+$ is the largest root
of $\D_r=0$, and its area is
\be
A=\frac{2\p^2(r_+^2+a^2)(r_+^2+b^2)}{r_+\X_a\X_b}.
\ee
The inverse temperature is given by
\be
\b=2\p\left[r_+\left(1+\frac{r_+^2}{l^2}\right)\left(\frac{1}{r_+^2+a^2}+
\frac{1}{r_+^2+b^2}\right)-\frac{1}{r_+}\right]^{-1}.
\ee
The angular velocities relative to a non-rotating frame at infinity are
\be
\Om_a=\frac{a(1+r_+^2l^{-2})}{r_+^2+a^2},\quad 
\Om_b=\frac{b(1+r_+^2l^{-2})}{r_+^2+b^2},
\ee
and the corresponding angular momenta are easily evaluated
\bea
&&J_a=\int_{\pa\cm\cap C}{\bf Q}[\pa_\f]=\frac{4\p^2ma}{\k^2\Xi_a^2\Xi_b},\\
&&J_b=\int_{\pa\cm\cap C}{\bf Q}[\pa_\psi]=\frac{4\p^2mb}{\k^2\Xi_a\Xi_b^2}.
\eea

As for the four dimensional Kerr-Newman-AdS black hole, in order to
bring the metric into the standard asymptotic form, we need to introduce 
the new coordinates

\bea
&&r=\bar{r}\left\{1+\frac14\hat{\D}_{\bar{\th}}\frac{l^2}{\bar{r}^2}+
\frac{1}{16}\hat{\D}_{\bar{\th}}(1+\hat{\X}_a+\hat{\X}_b-2\hat{\D}_{\bar{\th}})
\frac{l^4}{\bar{r}^4}+\co\left(\frac{1}{\bar{r}^6}\right)\right\},\\
&&\th=\bar{\th}+\frac{1}{16}(1-\hat{\D}_{\bar{\th}})\hat{\D}_{\bar{\th}}'
\frac{l^4}{\bar{r}^4}
-\frac{1}{32}(1-\hat{\D}_{\bar{\th}})\hat{\D}_{\bar{\th}}'(1+\hat{\X}_a+
\hat{\X}_b+3\hat{\D}_{\bar{\th}})
\frac{l^6}{\bar{r}^6}+\co\left(\frac{1}{\bar{r}^8}\right),\NO
\eea
where, to simplify the notation, we have defined
\be
\hat{\D}_\th=1-\D_\th,\quad \hat{\X}_a=1-\X_a,\quad \hat{\X}_b=1-\X_b.
\ee
In the new coordinate system the induced metric, up to 
terms of order $1/\bar{r}^6$ inside the braces, takes the form
\bea
&&\g_{\bar{\th}\bar{\th}}=\frac{\bar{r}^2}{\D_{\bar{\th}}}\left\{1+
\frac{3\hat{\D}_{\bar{\th}}}{2}\frac{l^2}{\bar{r}^2}+\frac14\left[(1-
\frac{3\hat{\D}_{\bar{\th}}}{2})(\hat{\X}_a+\hat{\X}_b-
\frac{3\hat{\D}_{\bar{\th}}}{2})+\hat{\X}_a\hat{\X}_b\right]\frac{l^4}
{\bar{r}^4}
\right\},\NO\\
&&\g_{tt}=-\frac{\bar{r}^2}{l^2}\left\{1+(1+\hat{\X}_a+
\hat{\X}_b-\frac{\hat{\D}_{\bar{\th}}}{2})\frac{l^2}{\bar{r}^2}+
\left[\frac{\hat{\D}_{\bar{\th}}}{8}(1+\hat{\X}_a+\hat{\X}_b
-\frac{3\hat{\D}_{\bar{\th}}}{2})-\frac{2m}{l^2}
\right]\frac{l^4}{\bar{r}^4}
\right\},\NO\\
\NO\\
&&\g_{t\f}=\frac{\bar{r}^2}{l^2}\frac{a\sin^2\bar{\th}}{\X_a}\left\{1+
(\hat{\X}_a+\frac{\hat{\D}_{\bar{\th}}}{2})\frac{l^2}{\bar{r}^2}+
\frac14\left[(\hat{\X}_b-\frac{\hat{\D}_{\bar{\th}}}{2})
(\X_a-\frac{\hat{\D}_{\bar{\th}}}{2})+
\hat{\X}_a\hat{\X}_b-\frac{8m}{l^2}\right]
\frac{l^4}{\bar{r}^4}
\right\},\NO\\\NO\\
&&\g_{t\psi}=\frac{\bar{r}^2}{l^2}\frac{b\cos^2\bar{\th}}{\X_b}\left\{1+
(\hat{\X}_b+\frac{\hat{\D}_{\bar{\th}}}{2})\frac{l^2}{\bar{r}^2}+
\frac14\left[(\hat{\X}_a-\frac{\hat{\D}_{\bar{\th}}}{2})
(\X_b-\frac{\hat{\D}_{\bar{\th}}}{2})+
\hat{\X}_a\hat{\X}_b-\frac{8m}{l^2}\right]
\frac{l^4}{\bar{r}^4}
\right\},\NO\\\NO\\
&&\g_{\f\f}=\bar{r}^2\frac{\sin^2\bar{\th}}{\X_a}\left\{1+
(\hat{\X}_a+\frac{\hat{\D}_{\bar{\th}}}{2})\frac{l^2}{\bar{r}^2}+
\frac14\left[(\hat{\X}_b-\frac{\hat{\D}_{\bar{\th}}}{2})
(\X_a-\frac{\hat{\D}_{\bar{\th}}}{2})+
\hat{\X}_a\hat{\X}_b+\frac{8m}{l^2}\frac{a^2\sin^2\bar{\th}}{l^2\X_a}\right]
\frac{l^4}{\bar{r}^4}
\right\},\NO\\\NO\\
&&\g_{\psi\psi}=\bar{r}^2\frac{\cos^2\bar{\th}}{\X_b}\left\{1+
(\hat{\X}_b+\frac{\hat{\D}_{\bar{\th}}}{2})\frac{l^2}{\bar{r}^2}+
\frac14\left[(\hat{\X}_a-\frac{\hat{\D}_{\bar{\th}}}{2})
(\X_b-\frac{\hat{\D}_{\bar{\th}}}{2})+
\hat{\X}_a\hat{\X}_b+\frac{8m}{l^2}\frac{b^2\cos^2\bar{\th}}{l^2\X_b}\right]
\frac{l^4}{\bar{r}^4}
\right\},\NO\\\NO\\
&&\g_{\f\psi}=\bar{r}^2\left\{\frac{2m}{l^2}\frac{a\cos^2\bar{\th}}{l\Xi_a}
\frac{b\sin^2\bar{\th}}{l\X_b}\frac{l^4}{\bar{r}^4}
\right\},
\eea
while the canonical radial coordinate $r_*$ is given by
\bea
&&dr_*=l\left\{1-\frac12(1+\hat{\X}_a+\hat{\X}_b)\frac{l^2}{\bar{r}^2}
+\left[\frac{m}{l^2}+\frac18(1+\hat{\X}_a+\hat{\X}_b)^2+
\frac14(1+\hat{\X}_a^2+\hat{\X}_b^2)\right]\frac{l^4}{\bar{r}^4}
\right\}\frac{d\bar{r}}{\bar{r}}.
\phantom{moreo}
\eea

\begin{flushleft}
{\em On-shell Euclidean action}
\end{flushleft}

The renormalized Euclidean action in five dimensions is given by 
\bea
I_{\rm ren} & = & -\frac{1}{2\k^2}\int_{\cm_{\bar{r}_o}}d^5x\sqrt{g_E}
\left(R[g_E]+\frac{12}{l^2}\right)\NO\\
&&-\frac{1}{\k^2}\int_{\S_{\bar{r}_o}}d^4x\sqrt{\g_E}\left(K
-\frac3l-\frac{l}{4}R+\frac{l^3}{16}(R_{ij}R^{ij}-\frac13R^2)\log 
e^{-2\bar{r}_o}\right).
\eea
Evaluating this expression we obtain
\be
I=\b M_{\rm Casimir}+
\frac{2\p^2\b}{\k^2 l^2\X_a\X_b}\left[ml^2-(r_+^2+a^2)(r_+^2+b^2)\right],
\ee
where
\be\label{Casimir_mass}
M_{\rm Casimir}\equiv
\frac{3\p^2 l^2}{4\k^2}\left(1+\frac{(\X_a-\X_b)^2}{9\X_a\X_b}\right).
\ee
This expression for the on-shell Euclidean action is precisely the expression
obtained in \cite{Awad&Johnson} and it differs from that
of \cite{GPP} by the term involving the Casimir
energy. Moreover, (\ref{Casimir_mass}) is equal  
the Casimir energy of the field theory on the rotating Einstein 
universe \cite{Awad&Johnson}.

Evaluating the holographic mass we find
\be
M \equiv \cq[\pa_t-\frac{a}{l^2}\pa_\f-\frac{b}{l^2}\pa_\psi]
= M_{\rm Casimir}+\frac{2\p^2 m(2\X_a+2\X_b-\X_a\X_b)}{\k^2\X_a^2\X_b^2},
\ee
which again agrees with the mass obtained in \cite{GPP}
except for the Casimir energy part. However, except for the Casimir 
energy, this mass is not the same as the one given in \cite{Awad&Johnson}.
The discrepancy arises presumably because \cite{Awad&Johnson} 
do not use the correct non-rotating timelike Killing vector 
to evaluate the mass. 

With the expressions for the mass and on-shell action we have obtained,
one can easily see that the quantum statistical relation (\ref{qm_relation})
is satisfied, despite the presence of the Casimir energy. However,
to show that our expressions do satisfy the first law, we need to examine the 
effect of an arbitrary variation of the parameters $a$, $b$ and $m$ 
on the representative of the conformal class at the boundary.

The boundary metric is
\be\label{b_metric} 
d\bar{s}^2=-dt^2+\frac{2a\sin^2\bar{\th}}{\X_a}dtd\f+
\frac{2b\cos^2\bar{\th}}{\X_b}dtd\psi+\frac{l^2}{\D_{\bar{\th}}}d\bar{\th}^2+
\frac{l^2\sin^2\bar{\th}}{\X_a}d\f^2+\frac{l^2\cos^2\bar{\th}}{\X_b}d\psi^2.
\ee
Under a variation of the angular parameters $a,\; b$, this metric 
is {\em not} kept fixed as is required by the variational problem.
The conformal class however is kept fixed (up to a diffeomorphism). To see 
this first consider the variation of (\ref{b_metric}) w.r.t. $a$ and $b$, and 
then perform the compensating infinitesimal diffeomorphism
\be\label{compensating_diffeo}
t=t',\quad
\tan^2\bar{\th}=\left(1+\frac{\d\X_a}{\X_a}-\frac{\d\X_b}{\X_b}\right)\tan^2
\bar{\th}',
\quad \f=\f'-\frac{\d a}{l^2}t',\quad \psi=\psi'-\frac{\d b}{l^2}t'.
\ee
The result of the combined transformation is
\be
d\bar{s}^2\to\left(1-\frac{\d\X_a}{\X_a}\sin^2\bar{\th}-\frac{\d\X_b}{\X_b}
\cos^2\bar{\th}\right) d\bar{s}^2.
\ee
The variation of the on-shell action due to this Weyl factor is
\be
\d_\s I=-\int_{\pa\cm}d^dx\sqrt{\g_E}\ca\d\s=\frac{\p^2\b l^2}{12\k^2}\d
\left(\frac{\X_a}{\X_b}+\frac{\X_b}{\X_a}\right)=\b\d M_{\rm Casimir}=
\b\d_\s M,
\ee
where the last equality follows from (\ref{Weyl}). Therefore, as expected,
only the Casimir energy part of the mass transforms non trivially under a Weyl 
transformation. 

Summarizing, we have shown that under a generic variation of the parameters
$a$, $b$ and $m$
\be
\d M=\d M_{\rm Casimir}+T\d S+\Om_a\d J_a+\Om_b\d J_b,
\ee
in complete agreement with (\ref{first_law_modified}). 
The first law then is satisfied once we accompany such a generic
variation with a compensating PBH transformation which undoes the
Weyl transformation of the representative of the conformal 
class.\footnote{Of course we should also perform a compensating 
diffeomorphism (\ref{compensating_diffeo}), but this does not 
affect the first law since all thermodynamic variables are
invariant under such a diffeomorphism.}

\section{Conclusion} 

We discussed in this paper the variational problem for AdS gravity,
the definition of conserved charges and the first law of thermodynamics
for asymptotically locally AdS black hole spacetimes. We conclude
by summarizing the main points.

AlAdS spacetimes are solutions of Einstein's equations whose
Riemann tensor is asymptotically equal to the Riemann
tensor of  $AdS$ but their asymptotic global
structure is not necessarily that of $AdS$. 
Their metric tensor
induces a conformal structure at infinity and so, a natural set of 
Dirichlet boundary conditions for AdS gravity is that a conformal 
structure is kept 
fixed. Notice that any other choice of Dirichlet 
boundary conditions would break part of the bulk diffeomorphisms,
namely the ones that induce a Weyl transformation at the boundary.
We examined the variational problem for such Dirichlet 
boundary conditions and found that it is well-posed provided
the conformal anomaly $\ca$ is zero and we add to the action
(in addition to the Gibbons-Hawking term) a set of new boundary terms.
These new boundary terms are precisely the boundary 
counterterms introduced in 
\cite{HS,Balasubramanian&Kraus} 
in order to achieve finiteness of the on-shell action and of the 
holographic stress energy tensor. If the conformal anomaly 
is non-zero, however, one has to choose a specific representative
of the boundary conformal structure to make the 
variational problem well-posed, thus breaking
part of the bulk diffeomorphisms. 
In this case the boundary counterterms guarantee that the on-shell
action has a well-defined transformation under the  
broken diffeomorphisms, the transformation rule being determined 
by the conformal anomaly. In other words, we need to pick 
a reference representative in this case, but the charge from 
one representative to another is essentially determined by the conformal 
class of the boundary metric via the conformal anomaly.

We then derived the conserved charges for AlAdS spacetimes that 
possess asymptotic symmetries. The holographic charges were 
originally derived \cite{Balasubramanian&Kraus,dHSS,Skenderis_proceedings}
using the Brown-York prescription \cite{Brown:1992br} supplemented
by appropriate boundary counterterms \cite{HS}. Here we derived
the conserved charges using Noether's method
and the covariant phase space method of Wald {\it et al} 
\cite{Wald93,Iyer:1994ys,Wald&Zoupas}
and found that they are equal to the
holographic charges.\footnote{The fact that the holographic charges 
are associated with asymptotic symmetries was also recently 
shown  in \cite{Hollands:2005ya} using somewhat different methods.} 
Notice that the case
of  $AAdS_{2k+1}$ spacetimes,
i.e. ones that approach asymptotically 
the exact $AdS_{2k+1}$ solution, is special in that there exists 
a covariantly conserved stress energy tensor constructed locally from 
the boundary metric that can be used to off-set the charges 
such that their value is equal to zero for the $AdS_{2k+1}$ solution.
This tensor is equal to the holographic stress 
energy tensor of $AdS_{2k+1}$ \cite{dHSS}. When such off-set is
done, i.e. when the charges are measured relative to $AdS_{2k+1}$,
the conserved charges agree 
with those of \cite{AM,Henneaux:1984xu}. A detailed comparison 
between different notions of conserved charges for AAdS spacetimes
was recently presented in \cite{Hollands:2005wt}.

We next considered general stationary, axisymmetric, charged AlAdS black 
holes in any dimension and showed in general that the quantum statistical
relation (or Smarr formula) and the first law of thermodynamics hold.
We would like to emphasize that
the variations that enter in the first law need not respect any of the
symmetries of the solution but they have to respect the boundary 
conditions. In other words, there are general normalizable variations
keeping fixed the non-normalizable mode (see footnote \ref{var_norm}). 
In some cases, such as
the five dimensional Kerr-AdS solution in Boyer-Lindquist coordinates,
the solution is parameterized such that the mass and other conserved charges
depend on parameters that also appear in the boundary metric. When 
varying these parameters one varies not only the conserved charges
but also the boundary metric, thus violating the boundary conditions. 
To keep fixed the non-normalizable mode one must perform a compensating 
coordinate transformation, and taking this into account one finds that 
the first law is satisfied, resolving a puzzle in the literature 
where it seemed that only the charges relative to $AdS$
satisfy the first law \cite{GKP}.

We illustrated our discussion by computing the conserved quantities
for the four dimensional Kerr-Newman-AdS and the five dimensional
Kerr-AdS black hole. An important point to realize is that 
the usual counterterms are defined
on the hypersurface $z=const.$, where $z$ is the Fefferman-Graham
coordinate. It is not in general correct to use the same set of 
counterterms when the cut-off hypersurface is different
(chosen for instance by considering $r=const.$ surfaces, where 
$r$ is a different radial coordinate that might appear naturally
in the bulk solution).  
So, to correctly compute the contribution of the counterterms to the on-shell
action one should asymptotically transform the bulk metric to
the Fefferman-Graham coordinates. Another subtle point is about the 
choice of timelike Killing vector in the definition of mass when the 
boundary metric is in a rotating frame. In this case one can resolve the 
ambiguity by choosing the timelike Killing vector that 
becomes the standard timelike Killing vector $\pa/\pa t$ in
a non-rotational frame.

\section*{Acknowledgments}

KS would like to thank Ecole Polytechnique for hospitality during the
completion of this work and for partial support from their grant
MEXT-CT-2003-509661. KS is funded by the NWO grant 016.023.007.

\begin{appendix}

\section*{Appendix}

\section{Gauge-fixed equations of motion}
\label{gauge_fixing}

In the neighborhood of the conformal boundary it is always possible to
write the bulk metric in the form
\beq\label{gf_metric}
ds^2=dr^2+\g_{ij}(r,x)dx^idx^j,
\eeq
where $r$ is a normal coordinate emanating from the boundary and
$\g_{ij}$ is the induced metric on the radial hypersurfaces $\S_r$. 
Choosing also the gauge $A_r=0$ for the gauge field, the gauge-fixed 
form of the equations of motion is

{\bf Einstein:}
\beqa\label{gf_einstein} 
&&K^2-K_{ij}K^{ij}=R+2\k^2 \tilde{T}_{d+1d+1},\nonumber \\
&&\label{einsteinIIgf}D_iK^i_j-D_jK=\k^2 \tilde{T}_{jd+1},\\
&&\dot{K}^i_j+KK^i_j=R^i_j-\k^2\left(\tilde{T}^i_j+\frac{1}
	{1-d}\tilde{T}^\s_\s\d^i_j\right).\nonumber
\eeqa
$\dot{K}^i_j$ here stands for $\frac{d}{dr}(\g^{ik}K_{kj})$ and 
$K_{ij}=\frac12\dot{\g}_{ij}$ is the extrinsic curvature of the hypersurfaces 
$\S_r$. Note also that the components of the Christoffel symbol of the bulk 
metric are
\beq\label{Christoffel} 
\G^{d+1}_{ij}=-K_{ij},\;\;\;\;\;\G^i_{d+1j}= K^i_j,
	\;\;\;\;\;\G^i_{jk}[g]=\G^i_{jk}[\g].
\eeq 

{\bf Vector:}
\bea\label{gf_vector}
&&D_i(U(\F)F^{ri})=0,\NO\\
&&\pa_r(U(\F)F^{rj})+KU(\F)F^{rj}+D_i(U(\F)F^{ij})=0.
\eea

{\bf Scalar:}
\bea\label{gf_scalar}
\pa_r(G_{IJ}(\F)\dot{\F}^J)+KG_{IJ}(\F)\dot{\F}^J+D^i
(G_{IJ}(\F)\pa_i\F^J)-\frac12\frac{\pa G_{JK}}{\pa\F^I}
(\dot{\F}^J\dot{\F}^K+\pa_i\F^J\pa^i\F^K)-\frac{\pa V}{\pa\F^I}\NO\\
-\frac14\frac{\pa U}{\pa \F^I}
(2\g^{ij}\dot{A}_i\dot{A}_j+F_{ij}F^{ij})=0.\NO\\
\eea

Here, $D_i$ is the covariant derivative with respect to the induced metric
$\g_{ij}$ and $F^r\phantom{}_i=\dot{A}_i$ in the gauge $A_r=0$.

\section{Asymptotic CKVs versus asymptotic bulk Killing vectors} 
\label{KV}

We discuss in this appendix the connection between asymptotic bulk isometries 
and boundary {\em conformal} isometries. In this discussion we will need a
well-known property of the linearized supergravity equations of motion, 
namely that for each bulk field they admit two linearly independent solutions, 
the {\em normalizable} and the {\em non-normalizable} modes, which 
near the boundary behave as $e^{-s_+ r}$ and   $e^{-s_- r}$  respectively.
The exponents $s_+,\,s_-$ are related to the scaling dimension of the
dual operators and the spacetime dimension. Specifically, we have
\bea
&& s^+  =  d-2,\quad s^-=-2, \quad {\rm for}\,\, \g_{ij},\NO\\
&& s^+  =  d-2,\quad s^-=0, \quad {\rm for}\,\, A_i,\NO\\
&& s^+  =  \D_I,\quad s^-=d-\D_I, \quad {\rm for}\,\, \F^I,
\eea
with $\D_I\geq d-\D_I$.

\begin{flushleft}
{\em Asymptotic conformal Killing vectors}
\end{flushleft}
{\it Definition:} We define an asymptotic conformal Killing vector (CKV) to be 
a bulk vector field $\xi$ which is asymptotically equal to 
a boundary conformal Killing vector. The precise asymptotic
conditions are
\be
(i)\quad\xi^r=\co(e^{-d r}),\quad (ii) \quad 
\xi^i(x,r) = \zeta^i(x) (1 + \co(e^{-(d+2)r}))
\ee 
where $\zeta^i(x)$ is a conformal Killing vector of 
$\mathtt{g}_{(0)}$.

The asymptotic conformal Killing vectors
are in one-to-one correspondence with asymptotic bulk Killing vectors, for 
if $\xi$ is an asymptotic CKV as defined above, then there exist 
$\hat{\xi}$, $\hat{\a}$, given in (\ref{PBH_param}) below, such that  
$\xi-\hat{\xi}$ is an asymptotic bulk Killing vector, up to a gauge 
transformation required to preserve the gauge fixing of the vector field, 
namely
\be 
\cl_{\xi-\hat{\xi}}\psi=\d_{\hat{\a}}\psi+\co(e^{-s_+ r}),
\ee 
or equivalently
\be \label{bkv}
\cl_\xi\psi=\cl_{\hat{\xi}}\psi+\d_{\hat{\a}}\psi+\co(e^{-s_+ r}).
\ee

To prove this we note that both $\cl_\xi \psi$ 
and $\cl_{\hat{\xi}}\psi+\d_{\hat{\a}}\psi$ satisfy the 
linearized equations of motion. As noted above, 
a basis for solutions of the the linearized equations 
of motion are the normalizable and non-normalizable solution.
Since in (\ref{bkv}) we require equality up to normalizable
mode, a sufficient condition for proving (\ref{bkv}) is that the 
leading asymptotics between the left and right hand side 
agree.  To show this we note that 
condition (i) and (\ref{lg})-(\ref{lf}) imply that in the gauge  
(\ref{gf_metric}) 
\be\label{BCKV}
\cl_\xi\psi=L_\xi\psi+\co(e^{-s_+r}).
\ee
Furthermore, condition (ii) is equivalent to
\be
L_\xi\psi=\frac1dD_i\xi^i\d_D\psi(1+ \co(e^{-r})).
\ee
It follows that the leading asymptotics agree with a 
PBH transformation with parameters,
\bea\label{PBH_param}
&&\hat{\xi}^r=\d\s(x),\NO\\
&&\hat{\xi}^i=\pa_j\d\s(x)\int_r^\infty dr'\g^{ji}(r',x),\NO\\
&&\hat{\a}=\pa_i\d\s(x)\int_r^\infty dr'A^i(r',x),
\eea
where
\be
\d\s=\frac1d D_i\xi^i,
\ee
which proves our assertion.

Notice that the asymptotic fall-off of $\xi^i$ in (ii) 
follows from the fact that
in order for a vector field to preserve the gauge (\ref{gf_metric})
we need
\be
\dot{\xi}^i=-\pa^i\xi^r \qquad \Rightarrow \qquad 
\dot{\xi}^i=\co(e^{-(d+2) r}).
\ee

\section{Proof of lemma 4.1}
\label{holographic=Noether}

In this appendix we give a proof of lemma \ref{hol=noether}.

\subsection*{Electric charge}

To prove (\ref{el_identity}) we start with the identity 
\be
\int_{\S_r\cap C}\ast\mathcal{F}=\int_{\S_r\cap C}d\s_i\frac{1}{\sqrt{-\g}}
\p^i,
\ee
where $\p^i=-\sqrt{-\g}U(\F)F^{ri}$ is the gauge field momentum. 
The second equation in (\ref{gf_vector}) can now be written as
\be\label{identity_1}
\dot{\p}^i=-\pa_j(\sqrt{-\g}U(\F)F^{ij}).
\ee
The momentum $\p^i$ and the radial derivative $\pa_r$ can be
expanded in eigenfunctions of the dilatation operator as in 
(\ref{momentum_exp}) and (\ref{radial_exp}) respectively. Moreover,
by Taylor expanding $U(\F)$ one obtains such an expansion for the RHS
of (\ref{identity_1}) too, which takes the form
\be
U(\F)F^{ij} = \left\{U(0)+\frac{\pa U}{\pa\F^I}\F^I+\frac{1}{2!}\frac{\pa^2 U}
{\pa\F^I\pa\F^J}\F^I\F^J+\cdots\right\}F^{ij}
 \equiv  \vf\sub{4}^{ij}+\vf\sub{5}^{ij}+\ldots.
\ee
Matching terms of the same dilatation weight on both sides of
(\ref{identity_1}) then we obtain 
\bea
{\p}\sub{3}^i=0,\NO\\
\sqrt{-\g}{\p}\sub{4}^i=-\frac{1}{d-4}\pa_j(\sqrt{-\g}\vf\sub{4}^{ij}),
\NO\\
\sqrt{-\g}{\p}\sub{5}^i=-\frac{1}{d-5}\pa_j\left[\sqrt{-\g}\vf\sub{5}^{ij}
-\frac{1}{d-4}\d\sub{1}\left(\sqrt{-\g}\vf\sub{4}^{ij}\right)\right],\NO\\
\vdots\phantom{more space here}\NO\\
\sqrt{-\g}\tilde{\p}\sub{d}^i=\frac{1}{2}\pa_j\left(\sqrt{-\g}\vf\sub{d}^{ij}
+\ldots\right).
\eea  
Therefore, all local terms in the momentum expansion are total derivatives 
while the non-local term ${\p}\sub{d}^i$ is left undetermined by this
iterative argument. Hence, 
\be\label{electric_charge}
\int_{\S_{r_o}\cap C}\ast\mathcal{F}=\int_{\S_{r_o}\cap C}d\s_i\frac{1}
{\sqrt{-\g}}\p^i=\int_{\S_{r_o}\cap C}d\s_i{\p}\sub{d}^i+\ldots.
\ee
Taking the limit $\S_{r_o}\to\pa\cm$ then completes the proof of 
(\ref{el_identity}).

\subsection*{Charges associated with asymptotic conformal isometries}

Applying a similar argument we now prove (\ref{CKV_identity}). Let, $\xi$ be 
an asymptotic conformal Killing vector as defined in appendix 
\ref{KV}, i.e.
\be
\cl_\xi\psi=\cl_{\hat{\xi}}\psi+\d_{\hat{\a}}\psi+\co(e^{-s_+r}),
\ee  
where $\hat{\xi}$ and $\hat{\a}$, given in (\ref{PBH_param}), generate a PBH 
transformation with conformal factor $\d\s=\frac1dD_i\xi^i$. Then, using
(\ref{Xi}), (\ref{B}) and the fact that in the gauge (\ref{gf_metric}) one
has
\bea\label{Xi_ri}
\Xi^{ri} & = & \nabla^{[r}\xi^{i]}+\k^2U(\F)F^{ri}A_j\xi^j\NO\\
& = & \dot{\xi}^i+\G_{rj}^i\xi^j-\frac{\k^2}{\sqrt{-\g}}\p^i A_j\xi^j\NO\\
& = & \left(K^i_j-\frac{\k^2}{\sqrt{-\g}}\p^i A_j\right)\xi^j+\co\left(
e^{-(d+2)r}\right),
\eea
we can write
\bea
\int_{\S_{r_o}\cap C} ({\bf Q}[\xi]-i_\xi{\bf B}) & = &
\frac{1}{\k^2}\int_{\S_{r_o}\cap C}d\s_i
\left(K^i_j-\frac{\k^2}{\sqrt{-\g}}\p^i A_j\right)\xi^j
-\frac{1}{\k^2}\int_{\S_{r_o}\cap C}d\s_i
\xi^i\left(K\sub{d}+\l_{\rm ct}\right)\NO\\\NO\\
&=&-\int_{\S_{r_o}\cap C}d\s_i\left[\left(2\pi\sub{d}^i_j+
\pi\sub{d}^i A_j\right)\xi^j+\co\left(
e^{-(d+2)r}\right)\right]\\
&&\phantom{morespacehere}+\frac{1}{\k^2}\int_{\S_{r_o}\cap C}d\s_i
\left(K^i_j-\frac{\k^2}{\sqrt{-\g}}\p^i A_j-\l\d^i_j\right)_{\rm ct}\xi^j.\NO
\eea
Taking the limit $\S_{r_o}\to\pa\cm$ we see that (\ref{CKV_identity}) is
equivalent to
\be\label{CKV_id}
\int_{\pa\cm\cap C}d\s_i\left(K^i_j-\frac{\k^2}{\sqrt{-\g}}\p^i A_j-
\l\d^i_j\right)_{\rm ct}\xi^j=0,
\ee
which we now prove.

From section \ref{Noether_charges} we know that on-shell  
\be\label{id_1}
{\rm d}{\bf Q}[\xi]+i_\xi{\bf L}={\bf \Theta}(\psi,\mathcal{L}_\xi\psi),
\ee
which, using (\ref{Theta}) and (\ref{Xi}), can be written as
\be
\nabla_\m\Xi^{\m\n}=\k^2\x^\n\left(-\cl_{\rm m}+\frac{1}{d-1}\tilde{T}^\s_\s
\right)-\k^2v^\n(\psi,\mathcal{L}_\xi\psi).
\ee 
In the gauge (\ref{gf_metric}) we can use (\ref{lambda_eq}) to get
\be
\pa_r\left[\sqrt{-\g}(\X^{ri}-\xi^i\l)\right]=\pa_j(\sqrt{-\g}\X^{ij})
-\k^2\sqrt{-\g}v^i(\psi,\mathcal{L}_\xi\psi)+\co\left(e^{-2 r}\right), 
\ee
or, using (\ref{Xi_ri}),
\be\label{CKV_relation}
\pa_r\left\{\left[\sqrt{-\g}\left(K^i_j-\l\d^i_j\right)-\k^2\p^i A_j\right]
\xi^j\right\}
=\pa_j(\sqrt{-\g}\X^{ij})-\k^2\sqrt{-\g}v^i(\psi,\mathcal{L}_\xi\psi)
+\co\left(e^{-2 r}\right). 
\ee 

To prove (\ref{CKV_id}) we only need the time component of this equation. 
In particular, if  $v^t(\psi,\mathcal{L}_\xi\psi)=\co(e^{-(d+2)r})$, then 
we can expand both sides of (\ref{CKV_relation}) in eigenfunctions of
the dilatation operator using (\ref{momentum_exp}), as was done for 
(\ref{identity_1}) in the previous section, and apply the same iterative 
argument to show that 
(\ref{CKV_id}) holds. Therefore, the proof of (\ref{CKV_identity}) is
complete once we show that  $v^t(\psi,\mathcal{L}_\xi\psi)=\co(e^{-(d+2)r})$.
As we now explain, this follows from (\ref{condition}).

From the explicit form of $v^t$, given in (\ref{v}), we see that
\be
v^t(\psi,\cl_\xi\psi)=v^t(\psi,\cl_{\hat{\xi}}\psi+\d_{\hat{\a}}\psi+
\co(e^{-s_+r}))=v^t(\psi,\cl_{\hat{\xi}}\psi+\d_{\hat{\a}}\psi)+
\co\left(e^{-(d+2)r}\right).
\ee
Moreover,
\bea
v^t(\psi,\cl_{\hat{\xi}}\psi+\d_{\hat{\a}}\psi) & = & -\frac{1}{2\k^2}
(\g^{t i}\g^{jk}-\g^{t k}\g^{ij})D_k\left(D_{(i}\hat{\xi}_{j)}+2K_{ij}\d\s
\right)\NO\\&&+U(\F)F^{t j}\left(L_{\hat{\x}}A_j+\dot{A}_j\d\s+\pa_j\hat{\a}
\right)+G_{IJ}(\F)\pa^t\F^I\left(\hat{\x}^i\pa_i\F^J+\dot{\F}^J\d\s\right)\NO
\\\NO\\
&=&-\frac{1}{2\k^2}
(\g^{t i}\g^{jk}-\g^{t k}\g^{ij})\left(D_kD_{(i}\hat{\xi}_{j)}+2K_{ij}D_k\d\s
\right)\NO\\&&+U(\F)F^{t j}\left(L_{\hat{\x}}A_j+\pa_j\hat{\a}\right)+
G_{IJ}(\F)\pa^t\F^I\hat{\x}^i\pa_i\F^J\\
&&-\frac{1}{\k^2}\left\{D^jK^t_j-D^t K-\k^2U(\F)F^{t j}\dot{A}_j
-\k^2G_{IJ}(\F)\pa^t\F^I\dot{\F}^J\right\}\d\s.\NO
\eea
The last term inside the braces vanishes by the second equation in 
(\ref{gf_einstein}).  From (\ref{condition}) and (\ref{PBH_param}) 
now follows that $\hat{\a}=0$ and $\hat{\xi}^i$ has no components along the 
isometry directions. Making repeated use of (\ref{condition}) it is then 
straightforward to show that 
$v^t(\psi,\cl_{\hat{\xi}}\psi+\d_{\hat{\a}}\psi)=0$, which completes the 
proof.

\section{Symplectic form on covariant phase space}
\label{symplectic}

In this appendix we give the explicit form of the symplectic current
on the covariant phase space as given by \cite{Crnkovic&Witten,Lee&Wald} 
(see also \cite{Wald&Zoupas}) and we
show that the corresponding pre-symplectic form is well-defined with
the boundary conditions (\ref{boundary_condition}), if there is no anomaly,
or (\ref{anbc}) when the anomaly is non-vanishing.
\begin{flushleft}
{\em Symplectic current}
\end{flushleft}
The symplectic current $D-1$-form is defined by \cite{Lee&Wald,Wald&Zoupas}
\be
{\bf \om}(\psi,\d_1\psi,\d_2\psi)=\d_2{\bf\Theta}(\psi,\d_1\psi)-
\d_1{\bf\Theta}(\psi,\d_2\psi).
\ee
The explicit form of this for the Lagrangian (\ref{Lagrangian}) can 
be derived directly from (\ref{v}). Writing  
\be
{\bf \om}(\psi,\d_1\psi,\d_2\psi)=-\ast{\bf \it w}(\psi,\d_1\psi,\d_2\psi),
\ee
with $ w^\m=w_{\rm gr}^\m+w_{\rm vec}^\m+w_{\rm sc}^\m$, we find
\bea\label{w_gr}\NO\\
w^\m_{\rm gr}&=&\frac{1}{2\k^2}\left(g^{\m\r}g^{\n\k}g^{\s\l}-\frac12
g^{\m\n}g^{\r\k}g^{\s\l}-\frac12 g^{\m\k}g^{\n\l}g^{\s\r}
-\frac12 g^{\m\r}g^{\n\s}g^{\k\l}+\frac12 g^{\m\n}g^{\r\s}g^{\k\l}\right)
\times \NO\\&&\left(\d_2g_{\k\l}\nabla_\n\d_1g_{\r\s}-\d_1g_{\k\l}\nabla_\n
\d_2g_{\r\s}\right),\\\NO\\\label{w_vec}
w^\m_{\rm vec} & = & U(\Phi)\left(\frac12 g^{\r\s}F^{\m\n}-g^{\m\s}F^{\r\n}-
g^{\n\s}F^{\m\r}\right)\left(\d_2g_{\r\s}\d_1A_\n-\d_1g_{\r\s}\d_2A_\n\right)
\NO\\
&&+\frac{\pa U(\Phi)}{\pa\Phi^I}F^{\m\n}\left(\d_1A_\n\d_2\Phi^I-\d_2A_\n\d_1
\Phi^I\right)\NO\\
&&+U(\Phi)(g^{\m\r}g^{\n\s}-g^{\m\s}g^{\n\r})(\d_1A_\n\nabla_\r\d_2A_\s-
\d_2A_\n\nabla_\r\d_1A_\s),\\\NO\\\label{w_sc}
w^\m_{\rm sc} & = & G_{IJ}(\Phi)\nabla_\r\Phi^J\left(\frac12 g^{\m\r}g^{\n\s}-
g^{\m\s}g^{\n\r}\right)(\d_2g_{\n\s}\d_1\Phi^I-\d_1g_{\n\s}\d_2\Phi^I)\NO\\
&&+\left(\frac{\pa G_{IJ}(\Phi)}{\pa\Phi^K}-\frac{\pa G_{KJ}(\Phi)}{\pa\Phi^I}
\right)\nabla^\m\Phi^J\d_1\Phi^I\d_2\Phi^K\NO\\
&&+G_{IJ}(\Phi)(\d_1\Phi^I\nabla^\m\d_2\Phi^J-\d_2\Phi^I\nabla^\m\d_1\Phi^J).
\\\NO
\eea

For the reader's convenience we now compile a list of the most important
properties of the symplectic current that we will need, along with the 
relevant proofs. Further details can be found in \cite{Lee&Wald, Wald&Zoupas}.

\begin{enumerate}

\item[I.] If $\psi$ satisfies the equations of motion and $\d_1\psi$, 
$\d_2\psi$ satisfy the linearized equations of motion, then $\om$ is closed
\be
{\rm d}{\bf \om}=0.
\ee

{\em Proof}:
Taking the second variation of (\ref{variation}) and using the fact that 
the functional derivatives of the Lagrangian commute we get
\bea
&&\d_2\d_1{\bf L}=\d_2{\bf E}\d_1\psi+{\rm d}\d_2{\bf \Theta}(\psi,\d_1\psi)=
\d_1{\bf E}\d_2\psi+{\rm d}\d_1{\bf \Theta}(\psi,\d_2\psi)=\d_1\d_2{\bf L}
\Rightarrow \NO\\
&&{\rm d}{\bf \om}(\psi,\d_1\psi,\d_2\psi)=\d_1{\bf E}\d_2\psi-\d_2{\bf E}
\d_1\psi.
\eea  
This completes the proof since $\d_1{\bf E}=\d_2{\bf E}=0$, by the hypothesis.

\item[II.] For an arbitrary fixed vector field $\xi$ on $\mathcal{M}$ and an
arbitrary gauge transformation $\a$, on-shell we have
\bea\label{diffeo_exact}
&&{\bf \om}(\psi,\d\psi,\mathcal{L}_\xi\psi)=
{\rm d}\left(\d{\bf Q}[\xi]-i_\xi{\bf \Theta}\right),\\
\label{gauge_exact}
&&{\bf \om}(\psi,\d\psi,\d_\a\psi)={\rm d}\d{\bf Q}_\a.
\eea

{\em Proof}: 
The variation of the diffeomorphism current with respect to an arbitrary 
variation $\d\psi$ of the fields (not necessarily satisfying the linearized 
equations of motion) is given by
\bea
\d{\bf J}[\xi] & = & \d{\bf\Theta}(\psi,\mathcal{L}_\xi\psi)-i_\xi\d{\bf L}\NO
\\
&=& \d{\bf\Theta}(\psi,\mathcal{L}_\xi\psi)-i_\xi {\rm d}{\bf \Theta}(\psi,\d
\psi)\NO
\\
&=& \d{\bf\Theta}(\psi,\mathcal{L}_\xi\psi)-\mathcal{L}_\xi{\bf\Theta}
(\psi,\d\psi)+{\rm d}(i_\xi{\bf\Theta}(\psi,\d\psi)),
\eea
where the equations of motion, ${\bf E}=0$, have been used together with
the identity $\mathcal{L}_\xi=i_\xi {\rm d}+{\rm d} i_\xi$ on forms. Since 
${\bf\Theta}$ 
is covariant with respect to bulk diffeomorphisms we have 
$\mathcal{L}_\xi{\bf\Theta}(\psi,\d\psi)=\d'{\bf\Theta}(\psi,\d\psi)$,
where $\d'\psi=\mathcal{L}_\xi\psi$. Hence,
\be
\d{\bf\Theta}(\psi,\mathcal{L}_\xi\psi)-\mathcal{L}_\xi{\bf\Theta}
(\psi,\d\psi)={\bf \om}(\psi,\d\psi,\mathcal{L}_\xi\psi),
\ee 
and so
\be\label{variation_id}
{\bf \om}(\psi,\d\psi,\mathcal{L}_\xi\psi)=
\d{\bf J}[\xi]-{\rm d}(i_\xi{\bf \Theta}).
\ee
Specializing this to solutions, $\d\psi$, of the linearized equations of motion
completes the proof of (\ref{diffeo_exact}). 

Moreover,
\be
{\bf \om}(\psi,\d\psi,\d_\a\psi)=\d{\bf\Theta}(\psi,\d_\a\psi)-
\d_\a{\bf\Theta}(\psi,\d\psi).
\ee
Gauge invariance implies that the second term on the RHS vanishes and
hence, on-shell, we obtain (\ref{gauge_exact}). 
\be
\ee

\item[III.] The pullback of the symplectic current on $\S_r$ takes the 
form
\bea\label{pullback_symplectic_form}
{\bf\om}(\psi,\d_1\psi,\d_2\psi) & = & \left\{\d_2(\sqrt{-\g}\p\sub{d}^{ij})
\d_1
\g_{ij}+\d_2(\sqrt{-\g}\p\sub{d}^i)\d_1 A _i+\d_2(\sqrt{-\g}\p\sub{\D_I}_I)
\d_1\F^I\right.\quad\quad\quad \NO\\&&\left. -1\leftrightarrow 2
\right\}d\m.
\eea

{\em Proof}: This follows immediately from the form of the
pullback (\ref{pullback_theta}) of ${\bf \Theta}$ on $\S_r$ together
with the commutativity of the field variations, $\d_2\d_1-\d_1\d_2=0$.

\end{enumerate}

\begin{flushleft}
{\em Pre-symplectic form}
\end{flushleft}

Having established the relevant properties of the symplectic current we now 
introduce the corresponding pre-symplectic 2-form on the field configuration 
space. Such a form induces a symplectic form on the solution submanifold 
of the configuration space \cite{Lee&Wald}. Given a Cauchy surface $C$, the 
pre-symplectic form relative to $C$ is defined  by \cite{Lee&Wald,Wald&Zoupas}
\be\label{symplectic_form}
\Om_C(\psi,\d_1\psi,\d_2\psi)=\int_C {\bf\om}(\psi,\d_1\psi,\d_2\psi).
\ee
In order for this to be well-defined obviously the integral
on the RHS of (\ref{symplectic_form}) must converge for all solutions $\psi$
of the field equations and any solutions $\d_1\psi,\,\,\d_2\psi$ of the
linearized equations of motion that satisfy the boundary conditions 
(\ref{boundary_condition}) - or (\ref{anbc}) in the case of non-vanishing
anomaly. These boundary conditions should also ensure that $\Om_C$ is
independent of the Cauchy surface $C$. 

To address these questions we note that the most general solution of 
the linearized equations of motion satisfying the boundary conditions 
(\ref{boundary_condition}) takes the form 
\be\label{linear_sol}
\d\psi=\cl_{\hat{\xi}}\psi+\d_{\hat{\a}}\psi+\hat{\d}\psi,
\ee
where $\hat{\xi}$, $\hat{\a}$, given by (\ref{PBH_param}), generate a
PBH transformation and $\hat{\d}\psi=\co(e^{-s^+ r})$
is an arbitrary {\em normalizable} solution. Since, as can be seen 
from (\ref{w_gr}), (\ref{w_vec}) and (\ref{w_sc}), the pullback of  
${\bf\om}(\psi,\d\psi,\hat{\d}\psi)$ onto $C$ is $\co(e^{-2r})$,
the only contribution to the pre-symplectic form which could be divergent
is the integral of ${\bf\om}(\psi,\cl_{\hat{\xi}_1}\psi+\d_{\hat{\a}_1}\psi,
\cl_{\hat{\xi}_2}\psi+\d_{\hat{\a}_2}\psi)$. However, if the background, 
$\psi$, satisfies the conditions of lemma \ref{hol=noether} and the Weyl 
factors $\d\s_1$ and $\d\s_2$ are independent of the coordinates adapted
to the isometries, then the 
pullback of ${\bf\om}(\psi,\cl_{\hat{\xi}_1}\psi+\d_{\hat{\a}_1}\psi,
\cl_{\hat{\xi}_2}\psi+\d_{\hat{\a}_2}\psi)$ onto the Cauchy surface $C$ 
vanishes. Hence, the defining integral (\ref{symplectic_form}) of $\Om_C$ is 
convergent.

Next, let $C$ and $C'$ be two Cauchy surfaces bounding a region 
$\D\subset \pa\cm$ of the boundary. Using Stokes' theorem and the fact 
that $\om$ is closed on-shell (property I), we get 
\be
\int_{C} {\bf\om}(\psi,\d_1\psi,\d_2\psi)-\int_{C'} {\bf\om}
(\psi,\d_1\psi,\d_2\psi)=\int_{\D\subset \pa\cm}{\bf\om}(\psi,\d_1\psi,
\d_2\psi).
\ee
Property III together with the boundary conditions (\ref{boundary_condition})
and the trace Ward identity (\ref{traceWI}) now give
\be\label{pullback_om_bc}
{\bf\om}(\psi,\d_1\psi,\d_2\psi)=\left\{\d_2(\sqrt{-\g}\mathcal{A})\d_1\s
-1\leftrightarrow 2\right\}d\m.
\ee
Therefore, $\Om_C$ is independent of the Cauchy surface provided we use
the boundary conditions  (\ref{boundary_condition}) when the anomaly vanishes,
and the boundary conditions (\ref{anbc}) when there is a non-zero anomaly.
This is in perfect agreement with our discussion of the variational problem.

\section{Electric part of the Weyl tensor and the Ashtekar-Magnon mass} 
\label{Weyl-tensor}

In this appendix we briefly discuss the connection between 
the `conformal mass' of \cite{AM} and  our analysis.
This issue is also discussed in the the recent work 
of \cite{Hollands:2005wt}. 

The authors of \cite{AM,AD} give a definition of the conserved charges 
for AAdS spacetimes in terms
of the electric part of the Weyl tensor, which, in the gauge (\ref{gf_metric}),
and for arbitrary matter fields, takes the form
\be
E^i_j=KK^i_j-K^i_kK^k_j-R^i_j+\frac{\k^2}{d-1}\left[(d-2)\tilde{T}^i_j
-\left((d-2)\frac1d\tilde{T}^\s_\s+\tilde{T}_{d+1 d+1}\right)\d^i_j\right].
\ee
This tensor is traceless due to the Hamilton constraint in (\ref{gf_einstein})
\be
E^i_i=K^2-K_{ij}K^{ij}-R-2\k^2 \tilde{T}_{d+1d+1}=0.
\ee

To make contact with their discussion let us specialize to pure gravity
in five dimensions (the inclusion of matter in the discussion  
is completely straightforward). Expanding this tensor in eigenfunctions of the
dilatation operator we immediately see that the term of weight 4 is given 
by
\be
E\sub{4}^i_j=2\left(K\sub{4}^i_j-K\sub{4}\d^i_j\right)+3K\sub{4}\d^i_j+
K\sub{2}K\sub{2}^i_j-K\sub{2}^i_kK\sub{2}^k_j.
\ee
Using now the expressions \cite{HS,dHSS,PS1}
\be
K\sub{2}^i_j=\frac12\left(R^i_j-\frac16R\d^i_j\right), \quad 
K\sub{4}=\frac{1}{24}\left(R^{ij}R_{ij}-\frac13R^2\right),
\ee
we obtain
\be
E\sub{4}^i_j=-2\k^2T\sub{4}^i_j+\frac14\left(-R^i_kR^k_j+\frac23RR^i_j+
\frac12R^k_lR^l_k\d^i_j-\frac14R^2\d^i_j\right),
\ee
where
\be
T\sub{4}^i_j\equiv -\frac{1}{\k^2}(K\sub{4}^i_j-K\sub{4}\d^i_j)
\ee
is the renormalized stress tensor. Therefore, in agreement with
Ashtekar and Das \cite{AM} and  Hollands, Ishibashi and Marolf 
\cite{Hollands:2005wt}, the difference between the holographic 
conserved charges, defined using $T\sub{4}^i_j$, and the 
Ashtekar-Magnon charges, defined using $E\sub{4}^i_j$, 
is the tensor
\be 
H^i_j\equiv \frac14\left(-R^i_kR^k_j+\frac23RR^i_j+
\frac12R^k_lR^l_k\d^i_j-\frac14R^2\d^i_j\right).
\ee
As discussed in the main text, this tensor is 
covariantly conserved and is equal to the 
holographic stress energy tensor of $AdS_5$ \cite{dHSS}.

There is a similar local tensor that 
is covariantly conserved when the metric is conformally flat
in all even dimensions: 
it is the holographic stress energy tensor of $AdS_{2k+1}$.
As it was shown in \cite{Skenderis:1999nb}, and reviewed
in section 2, see (\ref{AdScoef}), the Fefferman-Graham expansion 
of $AdS_{(2k+1)}$ terminates at order $z^4$ and all terms are locally
related to $\mathtt{g}_{(0)}$. It follows that the holographic
stress energy tensor, which in general contains 
the non-local (w.r.t. $\mathtt{g}_{(0)}$ ) term $\mathtt{g}_{(d)}$, 
is local in this case. The explicit expression for $d=6$ is 
given in (3.21) of \cite{dHSS}.

\end{appendix}


\begin{thebibliography}{99}

\bibitem{AM}
A. Ashtekar and A. Magnon, ``Asymptotically anti-de Sitter space-times'',
Class. Quant. Grav. {\bf 1} (1984) L39-L44;


\bibitem{Henneaux:1984xu}
M.~Henneaux and C.~Teitelboim,
``Hamiltonian Treatment Of Asymptotically Anti-De Sitter Spaces,''
Phys.\ Lett.\ B {\bf 142} (1984) 355;
``Asymptotically Anti-De Sitter Spaces,''
Commun.\ Math.\ Phys.\  {\bf 98}, 391 (1985).

\bibitem{FG}
C. Fefferman and C. Robin Graham, ``Conformal Invariants'', in 
{\it Elie Cartan et les Math\'ematiques d'aujourd'hui} (Ast\'erisque, 1985) 
95.

\bibitem{Graham} C.R. Graham, ``Volume and Area Renormalizations for 
Conformally Compact Einstein Metrics'', math.DG/9909042.


\bibitem{anderson}
  M.~T.~Anderson,
  ``Geometric aspects of the AdS/CFT correspondence,''
  arXiv:hep-th/0403087.

\bibitem{Abbott:1981ff}
  L.~F.~Abbott and S.~Deser,
  ``Stability Of Gravity With A Cosmological Constant,''
  Nucl.\ Phys.\ B {\bf 195} (1982) 76.

\bibitem{Hawking:1995fd}
  S.~W.~Hawking and G.~T.~Horowitz,
  ``The Gravitational Hamiltonian, action, entropy and surface terms,''
  Class.\ Quant.\ Grav.\  {\bf 13} (1996) 1487
  [arXiv:gr-qc/9501014].

\bibitem{Chrusciel:2001qr}
  P.~T.~Chrusciel and G.~Nagy,
  ``The mass of spacelike hypersurfaces in asymptotically anti-de Sitter
  space-times,''
  Adv.\ Theor.\ Math.\ Phys.\  {\bf 5} (2002) 697
  [arXiv:gr-qc/0110014].


\bibitem{Hollands:2005wt}
  S.~Hollands, A.~Ishibashi and D.~Marolf,
  ``Comparison between various notions of conserved charges in asymptotically
  AdS-spacetimes,''
  arXiv:hep-th/0503045.


\bibitem{Magnon:1985sc}
  A.~Magnon,
  ``On Komar Integrals In Asymptotically Anti-De Sitter Space-Times,''
  J.\ Math.\ Phys.\  {\bf 26}, 3112 (1985).


\bibitem{Deruelle}
  N.~Deruelle and J.~Katz,
  ``On the mass of a Kerr-anti-de Sitter spacetime in D dimensions,''
  Class.\ Quant.\ Grav.\  {\bf 22} (2005) 421
  [arXiv:gr-qc/0410135];
  N.~Deruelle,
  ``Mass and angular momenta of Kerr-anti-de Sitter spacetimes,''
  arXiv:gr-qc/0502072.


\bibitem{Barnich&Compere}
  G.~Barnich and G.~Compere,
  ``Generalized Smarr relation for Kerr AdS black holes from improved surface
  integrals,''
  Phys.\ Rev.\ D {\bf 71} (2005) 044016
  [arXiv:gr-qc/0412029].


\bibitem{ads/cft}
J.~M.~Maldacena,
  ``The large N limit of superconformal field theories and supergravity,''
  Adv.\ Theor.\ Math.\ Phys.\  {\bf 2} (1998) 231
  [Int.\ J.\ Theor.\ Phys.\  {\bf 38} (1999) 1113]
  [arXiv:hep-th/9711200],

\bibitem{GKP}
  S.~S.~Gubser, I.~R.~Klebanov and A.~M.~Polyakov,
  ``Gauge theory correlators from non-critical string theory,''
  Phys.\ Lett.\ B {\bf 428} (1998) 105
  [arXiv:hep-th/9802109].

\bibitem{Witten:1998qj}
E.~Witten,
``Anti-de Sitter space and holography,''
Adv.\ Theor.\ Math.\ Phys.\  {\bf 2}, 253 (1998)
[hep-th/9802150].


\bibitem{HS}
M.~Henningson and K.~Skenderis,
``The holographic Weyl anomaly,''
JHEP {\bf 9807} (1998) 023
[hep-th/9806087];
M.~Henningson and K.~Skenderis,
``Holography and the Weyl anomaly,''
Fortsch.\ Phys.\  {\bf 48} (2000) 125
[hep-th/9812032].

\bibitem{Balasubramanian&Kraus}
  V.~Balasubramanian and P.~Kraus,
  ``A stress tensor for anti-de Sitter gravity,''
  Commun.\ Math.\ Phys.\  {\bf 208} (1999) 413
  [arXiv:hep-th/9902121].

\bibitem{dHSS}
S.~de Haro, S.~N.~Solodukhin and K.~Skenderis,
``Holographic reconstruction of spacetime and renormalization in the
 AdS/CFT correspondence,''
Commun.\ Math.\ Phys.\  {\bf 217} (2001) 595
[hep-th/0002230].

\bibitem{Skenderis_proceedings}
K.~Skenderis,
``Asymptotically anti-de Sitter spacetimes and their stress energy  tensor,''
Int.\ J.\ Mod.\ Phys.\ A {\bf 16}, 740 (2001)
[arXiv:hep-th/0010138].

\bibitem{Brown:1992br}
  J.~D.~Brown and J.~W.~.~York,
  ``Quasilocal energy and conserved charges derived from the gravitational
  action,''
  Phys.\ Rev.\ D {\bf 47}, 1407 (1993).


\bibitem{GPP}
  G.~W.~Gibbons, M.~J.~Perry and C.~N.~Pope,
  ``The first law of thermodynamics for Kerr - anti-de Sitter black holes,''
  arXiv:hep-th/0408217.


\bibitem{Hollands:2005ya}
  S.~Hollands, A.~Ishibashi and D.~Marolf,
  ``Counter-term charges generate bulk symmetries,''
  arXiv:hep-th/0503105.

\bibitem{Lee&Wald}
  J.~Lee and R.~M.~Wald,
  ``Local Symmetries And Constraints,''
  J.\ Math.\ Phys.\  {\bf 31}, 725 (1990).

\bibitem{Wald&Zoupas}
  R.~M.~Wald and A.~Zoupas,
  ``A General Definition of "Conserved Quantities" in General Relativity and
  Other Theories of Gravity,''
  Phys.\ Rev.\ D {\bf 61}, 084027 (2000)
  [arXiv:gr-qc/9911095].

\bibitem{Skenderis:2002wp}
K.~Skenderis,
``Lecture notes on holographic renormalization,''
Class.\ Quant.\ Grav.\  {\bf 19} (2002) 5849
[hep-th/0209067].

\bibitem{penrose}
R. Penrose and W. Rindler, {\it Spinors and Spacetime}, vol. 2 (Cambridge 
University) chapter 9.


\bibitem{Taylor-Robinson:2000xw}
  M.~M.~Taylor-Robinson,
  ``More on counterterms in the gravitational action and anomalies,''
  arXiv:hep-th/0002125.

\bibitem{holren}
M.~Bianchi, D.~Z.~Freedman and K.~Skenderis,
``How to go with an RG flow,''
JHEP {\bf 0108} (2001) 041
[arXiv:hep-th/0105276];
``Holographic renormalization,''
Nucl.\ Phys.\ B {\bf 631} (2002) 159
[arXiv:hep-th/0112119].


\bibitem{PS1}
I.~Papadimitriou and K.~Skenderis,
``AdS/CFT correspondence and geometry,''
[hep-th/0404176].


\bibitem{PS2}
  I.~Papadimitriou and K.~Skenderis,
  ``Correlation functions in holographic RG flows,''
  JHEP {\bf 0410} (2004) 075
  [arXiv:hep-th/0407071].

\bibitem{Skenderis:1999nb}
K.~Skenderis and S.~N.~Solodukhin,
``Quantum effective action from the AdS/CFT correspondence,''
Phys.\ Lett.\ B {\bf 472} (2000) 316
[arXiv:hep-th/9910023].

\bibitem{PS}
A.~Petkou and K.~Skenderis,
``A non-renormalization theorem for conformal anomalies,''
Nucl.\ Phys.\ B {\bf 561} (1999) 100
[arXiv:hep-th/9906030].


\bibitem{Kraus:1999di}
P.~Kraus, F.~Larsen and R.~Siebelink,
``The gravitational action in asymptotically AdS and flat spacetimes,''
Nucl.\ Phys.\ B {\bf 563} (1999) 259
[hep-th/9906127].


\bibitem{deBoer:1999xf}
J.~de Boer, E.~Verlinde and H.~Verlinde,
``On the holographic renormalization group,''
JHEP {\bf 0008} (2000) 003
[hep-th/9912012];
J.~de Boer,
``The holographic renormalization group,''
Fortsch.\ Phys.\  {\bf 49}, 339 (2001)
[hep-th/0101026].

\bibitem{Martelli:2002sp}
D.~Martelli and W.~Muck,
``Holographic renormalization and Ward identities with the  
Hamilton-Jacobi method,''
Nucl.\ Phys.\ B {\bf 654} (2003) 248
[arXiv:hep-th/0205061];

\bibitem{Brown:1986nw}
  J.~D.~Brown and M.~Henneaux,
  ``Central Charges In The Canonical Realization Of Asymptotic Symmetries: An
  Example From Three-Dimensional Gravity,''
  Commun.\ Math.\ Phys.\  {\bf 104} (1986) 207.

\bibitem{Imbimbo:1999bj}
  C.~Imbimbo, A.~Schwimmer, S.~Theisen and S.~Yankielowicz,
  ``Diffeomorphisms and holographic anomalies,''
  Class.\ Quant.\ Grav.\  {\bf 17}, 1129 (2000)
  [arXiv:hep-th/9910267].


\bibitem{GH}
  G.~W.~Gibbons and S.~W.~Hawking,
  ``Action Integrals And Partition Functions In Quantum Gravity,''
  Phys.\ Rev.\ D {\bf 15} (1977) 2752.

\bibitem{Wald_book}
R. M. Wald, {\em General Relativity}, The University of Chicago Press 1984.

\bibitem{Carter69}
B.~Carter, 
``Killing Horizons and Orthogonally Transitive Groups in Space-Time,''
J.\ Math.\ Phys. {\bf 10} (1969) 70-81.

\bibitem{Heusler}
  M.~Heusler,
  ``No-hair theorems and black holes with hair,''
  Helv.\ Phys.\ Acta {\bf 69} (1996) 501
  [arXiv:gr-qc/9610019].

\bibitem{Emparan&Reall}
  R.~Emparan and H.~S.~Reall,
  ``Generalized Weyl solutions,''
  Phys.\ Rev.\ D {\bf 65} (2002) 084025
  [arXiv:hep-th/0110258].


\bibitem{Harmark}
  T.~Harmark,
  ``Stationary and axisymmetric solutions of higher-dimensional general
  relativity,''
  Phys.\ Rev.\ D {\bf 70} (2004) 124002
  [arXiv:hep-th/0408141].



\bibitem{Hawking:1998ct}
  S.~W.~Hawking, C.~J.~Hunter and D.~N.~Page,
  ``Nut charge, anti-de Sitter space and entropy,''
  Phys.\ Rev.\ D {\bf 59} (1999) 044033
  [arXiv:hep-th/9809035].

\bibitem{Awad:2000gg}
  A.~Awad and A.~Chamblin,
  ``A bestiary of higher dimensional Taub-NUT-AdS spacetimes,''
  Class.\ Quant.\ Grav.\  {\bf 19} (2002) 2051
  [arXiv:hep-th/0012240].

\bibitem{Olea:2005gb}
  R.~Olea,
  ``Mass, angular momentum and thermodynamics in four-dimensional Kerr-AdS
  black holes,''
  arXiv:hep-th/0504233;
P.~Mora, R.~Olea, R.~Troncoso and J.~Zanelli,
  ``Vacuum energy in odd-dimensional AdS gravity,''
  arXiv:hep-th/0412046.


\bibitem{Crnkovic&Witten}
C.~Crnkovi\'c and E.~Witten,
``Covariant description of canonical formalism in geometrical theories''
in {\em Three Hundred Years of Gravitation}, edited by S.~W.~Hawking and
W.~Israel (Cambridge U.P., Cambridge, 1987).

\bibitem{zuckerman} G.J. Zuckerman, 
``Action Principles and Global Geometry,''
in {\em Mathematical Aspects of String Theory},
San Diego 1986, Proceedings, Ed. S.-T. Yau
(World Scientific, 1987). p. 259.

\bibitem{BD} N.D. Birrell and P.C.W. Davies, 
``Quantum fields in curved space'', Cambridge Monographs on
Mathematical Physics, chapter 6.

\bibitem{AD}
A.~Ashtekar and S.~Das,
  ``Asymptotically anti-de Sitter space-times: Conserved quantities,''
  Class.\ Quant.\ Grav.\  {\bf 17} (2000) L17
  [arXiv:hep-th/9911230].

\bibitem{Wald93}
  R.~M.~Wald,
  ``Black hole entropy is the Noether charge,''
  Phys.\ Rev.\ D {\bf 48} (1993) 3427
  [arXiv:gr-qc/9307038].

\bibitem{Jacobson:1993vj}
  T.~Jacobson, G.~Kang and R.~C.~Myers,
  ``On black hole entropy,''
  Phys.\ Rev.\ D {\bf 49}, 6587 (1994)
  [arXiv:gr-qc/9312023].

\bibitem{Iyer:1994ys}
  V.~Iyer and R.~M.~Wald,
  ``Some properties of Noether charge and a proposal for dynamical black hole
  entropy,''
  Phys.\ Rev.\ D {\bf 50}, 846 (1994)
  [arXiv:gr-qc/9403028].


\bibitem{Kostelecky:1995ei}
  V.~A.~Kostelecky and M.~J.~Perry,
  ``Solitonic Black Holes in Gauged N=2 Supergravity,''
  Phys.\ Lett.\ B {\bf 371} (1996) 191
  [arXiv:hep-th/9512222].

\bibitem{HHTR}
  S.~W.~Hawking, C.~J.~Hunter and M.~M.~Taylor-Robinson,
  ``Rotation and the AdS/CFT correspondence,''
  Phys.\ Rev.\ D {\bf 59} (1999) 064005
  [arXiv:hep-th/9811056].



\bibitem{Das&Mann}
  S.~Das and R.~B.~Mann,
  ``Conserved quantities in Kerr-anti-de Sitter spacetimes in various
  dimensions,''
  JHEP {\bf 0008} (2000) 033
  [arXiv:hep-th/0008028].

\bibitem{Silva:2002jq}
  S.~Silva,
  ``Black hole entropy and thermodynamics from symmetries,''
  Class.\ Quant.\ Grav.\  {\bf 19} (2002) 3947
  [arXiv:hep-th/0204179].

\bibitem{Caldarelli}
  M.~M.~Caldarelli, G.~Cognola and D.~Klemm,
  ``Thermodynamics of Kerr-Newman-AdS black holes and conformal field
  theories,''
  Class.\ Quant.\ Grav.\  {\bf 17} (2000) 399
  [arXiv:hep-th/9908022].

\bibitem{Awad&Johnson}
  A.~M.~Awad and C.~V.~Johnson,
  ``Higher dimensional Kerr-AdS black holes and the AdS/CFT correspondence,''
  Phys.\ Rev.\ D {\bf 63} (2001) 124023
  [arXiv:hep-th/0008211].

\bibitem{Carter68}
B.~Carter, ``Hamilton-Jacobi and Schr\"odinger separable solutions 
of Einstein's equations,'' 
Commun.\ Math.\ Phys.{\bf 10} (1968) 280.

\bibitem{Plebanski} 
J.~F.~Plebanski and M.~Demianski,
``Rotating, charged, and uniformly accelerating mass in general 
relativity,'' 
Ann.\ Phys. {\bf 98} (1976) 98-127. 

\end{thebibliography}
\end{document}